\documentclass[aps,prc,reprint,showpacs,groupedaddress,onecolumn]{revtex4-1}

\usepackage{graphicx,color} 
\usepackage{dcolumn}  
\usepackage{bm}       
\usepackage{amsmath}
\usepackage{amssymb}
\usepackage{amsfonts}

\def\beq{\begin{equation}}
\def\eeq{\end{equation}}

\begin{document}

\title{Relativistic Formulation of Reaction Theory}

\author{W.~N.~Polyzou}
\affiliation{Department of Physics and Astronomy, The University of
Iowa, Iowa City, IA 52242, USA}

\author{Ch. Elster}
\affiliation{Institute of Nuclear and Particle Physics,  and
Department of Physics and Astronomy,  Ohio University, Athens, OH 45701,
USA}

\date{\today}

\begin{abstract}

A relativistic formulation of reaction theory for nuclei with a
dynamics given by a unitary representation of the Poincar\'e group is
developed.  Relativistic dynamics is introduced by starting from a
relativistic theory of free particles to which rotationally invariant
interactions are added to the invariant mass operator.  Poincar\'e
invariance is realized by requiring that simultaneous eigenstates of
the mass and spin transform as irreducible representations of the
Poincar\'e group.  A relativistic formulation of scattering theory is
presented and approximations emphasizing dominant degrees of freedom
that preserve unitarity, exact Poincar\'e invariance and exchange
symmetry are discussed.  A Poincar\'e invariant formulation of a (d,p)
reaction as a three-body problem is given as an explicit example.

\end{abstract}

\maketitle


\section{Introduction}
\label{intro}

The physics of exotic nuclei has become a major subject within
nuclear physics. A new generation of radioactive beam facilties such as
RIBF at RIKEN in Japan, FAIR at GSI in Germany, SPIRAL2 at GANIL in
France, and FRIB at MSU in the USA have been or will be soon in
operation. With the access to exotic nuclei at the limits of
nuclear stability, the physics of neutron and proton driplines has
become a focus of interest. Nuclei close to these driplines exhibit
phenomena different from the known stable ones, like the normal shell
closures may disappear and be replaced by new magic numbers, or
threshold phenomena like nuclear halo states may occur (for reviews see
e.g. \cite{Hagino-book,JonsonPRep}).

Ongoing and planned technical developments in beam production as well 
as in detection systems allow not only experiments with a larger variety
of nuclei, but more importantly allow measurements of reactions which were
traditionally carried out with stable beams like knock-out or transfer
reactions at a variety of energies. Even first polarization experiments
with radioactive beams are now possible~\cite{Sakaguchi:2013uut}.

Theoretical developments have been moving at a somewhat slower pace.
Here one should note, that at first the light exotic nuclei received
most of the theoretical attention, and approaches describing their
reactions were developed for a higher energy regime, where it was
believed that reactions are dominated by a few degrees of freedom and
thus approximations are justified.  These include the eikonal
approximation, or the adiabatic approximation in which degrees of
freedom are frozen. Many theoretical advances made over the last
decade however concentrate on the lower energy regime (lower than
roughly 50~MeV per nucleon) to energies relevant for astrophysical
processes.  This energy regime is the realm of non-relativistic
quantum mechanics, in which e.g. coupled discretized continuum channel
(CDCC) methods are applied to direct reactions. A review of selected
methods is given in Ref.~\cite{nunes-alkhalili}. Direct reactions also
lend themselves to adoptions of few-body techniques, which are well
established in the non-relativistic regime.  Well-defined examples
here are the (d,p) reaction on light nuclei, which can be successfully
described by a modified Faddeev approach~\cite{Deltuva:2009fp}.  In
fact, one of the experimentally as well as theoretically most
carefully studied system is the three-nucleon system, since the
Faddeev equations for three nucleons can be exactly solved for neutron
and proton-deuteron scattering, including two- and three-nucleon
forces~\cite{wgphysrep,Deltuva:2006ch}.

A Poincar\'e invariant formulation of the Faddeev equations was
pioneered in model calculations up to 1~GeV for spin-independent
forces~\cite{Lin:2007ck,Lin:2007kg,Lin:2008sy} and then employed for
realistic two and three-nucleon forces~\cite{Witala:2011yq} for
projectile energies up to 250~MeV. Both, the realistic as well as the
model calculation indicate that differential cross sections exhibit
some differences between a non-relativistic and relativistic treatment
at large momentum transfers, already as low as 250~MeV. In addition, those
calculations clearly indicate, that breakup reactions are considerably
more sensitive to a correct treatment of Poincar\'e invariance than
elastic scattering. The model calculation in Refs.~\cite{Lin:2007kg}
show that the correct Poincar\'e invariant treatment of
relativistic kinematics and dynamics in exclusive breakup reaction
cross sections can differ up to an order of magnitude from the Galilei
invariant calculation already at projectile energies around 500~MeV,
while at the same energy elastic scattering cross sections only differ
by about 15\%. In proton-deuteron elastic scattering using realistic
nucleon-nucleon (NN) forces the experimental cross section at back
angles is underpredicted even after enhancements due to both
three-body forces and relativistic effects are
included~\cite{Witala:2011yq,witala1}.  Given that back angles are
more sensitive to short distance physics, this suggests additional
degrees of freedom may be relevant~\cite{Platonova:2010zza}.

The purpose of this work is to develop a Poincar\'e invariant theory
of nuclear reactions in order to interpret experimental information
obtained in few GeV scale nuclear reactions.  While the formulation of
multiple scattering theories in a Galilei invariant framework has a
long tradition~\cite{watson,kmt}, Poincar\'e invariant counterparts do
not exist.  This article is a first step in this direction by
formulating a Poincar\'e invariant quantum theory of nuclear reactions 
that is dominated by a limited number of important degrees of freedom.

Relativistic invariance of a quantum theory requires the invariance of
quantum probabilities, expectation values and ensemble averages 
with respect to changes in the inertial coordinate
system.  This requires that the dynamics is given by a unitary
representation of the Poincar\'e group \cite{wigner:1939}.  Here the
Poincar\'e group refers to the transformations continuously connected
to the identity; invariance with respect to space reflections and time
reversal is not required and is not satisfied by the weak interaction.
Approximations that emphasize dominant degrees of freedom that
preserve both unitarity and exact Poincar\'e invariance are discussed
below.

The simplest way to construct a relativistic dynamics is to start with
a relativistic theory of free particles.  Interactions can be added in
a manner that preserves the overall Poincar\'e invariance.  We do this
in three steps.  First we boost the non-interacting $N$-body system to
the $N$-body rest frame.  Second, we add rotationally invariant
interactions to the non-interacting rest Hamiltonian, which in the
relativistic case is the $N$-particle invariant mass operator.  Third,
we solve for simultaneous eigenstates of the interacting mass and
spin, which can be done because of the rotational invariance of the
interaction.  The mass and spin are the invariant labels for irreducible
representations of the Poincar\'e group.  The relativistic dynamics is
defined by requiring that in all other inertial frames these mass-spin
eigenstate transform irreducibly with respect to the Poincar\'e group.
Once we have these operators the treatment of the reaction theory is similar
to the non-relativistic treatment.

In the first section we derive the transformation properties for a
single relativistic particle, then we consider the case of $N$
non-interacting relativistic particles.  We discuss how to decompose
products of irreducible representations of the Poincar\'e group into
direct integrals of irreducible representations.  In the third section
we add interactions to the mass Casimir operator of the
non-interacting irreducible representations to construct dynamical
unitary representations of the Poincar\'e group.  Then we focus on
reaction-theory models and relativistic scattering theory. After
addressing the treatment of identical particles, we explicitly
consider (d,p) reactions as an illustration of the general formulation.


\section{Relativistic Kinematics}
\label{kinematics}

In this section we discuss the first step, which is the relativistic
description of a single particle.  The state of a single particle of
mass $m$ and spin $j$ is characterized by its momentum, $\mathbf{p}$,
and the projection of its spin, $\mu$, on a given axis.  These are a
complete set of commuting observables for a structureless particle.
Simultaneous eigenstates of these observables, denoted by $\vert (m,j)
\mathbf{p},\mu \rangle$, are a basis for a single-particle Hilbert
space, ${\cal H}_{mj}$.

A unitary representation of the Poincar\'e group on ${\cal H}_{mj}$ 
is the product of a unitary representation, $U(\Lambda)$, of the Lorentz
group and a unitary space-time translation operator $T(a)$,
\beq
U(\Lambda ,a) = T(a) U(\Lambda) = U(\Lambda) T(\Lambda^{-1}a), 
\eeq
where $\Lambda$ is a Lorentz transformation and $a$ is a constant
four vector.  Explicit representations can
be constructed by considering the transformation properties of
rotations, $\Lambda = R$, translations, and Lorentz boosts, 
$\Lambda = B(\mathbf{p}/m)$, on rest
($\mathbf{0}$-momentum) eigenstates.

A particle at rest remains at rest under rotations.  On the other hand
the spins undergo rotations.  If the particle has spin $j$ then the
rest eigenstates transform under a $2j+1$ dimensional unitary
representation of the rotation group.  These elementary
transformations are
\beq
U(R) \vert (m,j)\mathbf{0},\mu \rangle
= \sum_{\nu=-j}^j\vert (m,j)\mathbf{0},\nu \rangle
D^j_{\nu \mu} (R) 
\label{r.1}
\eeq
where $D^j_{\nu \mu} (R)$ is an ordinary Wigner $D$-function, which is
a $2j+1$ dimensional unitary representation of the rotation group.

Since these states are rest eigenstates of the four-momentum, it also 
follows that under space-time translations by $a$, 
\beq
T(a) \vert (m,j)\mathbf{0},\mu \rangle =
e^{-i m a^0} \vert (m,j)\mathbf{0},\mu \rangle ,
\label{r.2}
\eeq
where $a^0$ is the $0$-component of $a$.

Because sequences of Lorentz boosts can generate rotations, we need an
unambiguous definition of a spin observable in frames moving with
momentum $\mathbf{p}$ relative to the rest frame.  There are many
possible definitions.

We define the spin observable in a general frame by the requirement
that it does not Wigner rotate when it is transformed to the
particle's rest frame by a rotationless Lorentz transformation
\beq
U(B (\mathbf{p}/m)) \vert (m,j)\mathbf{0},\mu \rangle :=
\vert (m,j)\mathbf{p},\mu \rangle
\sqrt{{\omega_m (\mathbf{p}) \over m}}.
\label{r.3}
\eeq
This is normally referred to as the canonical spin.

The rotationless Lorentz boost $B (\mathbf{p}/m)$ is the usual
textbook Lorentz boost that is normally expressed in terms of
hyperbolic sines and cosines of a rapidity, $\rho$.  The rotationless boost
from the particle's rest frame to a frame where it has momentum
$\mathbf{p}$ is
\beq
B (\mathbf{p}/m) := B (\mathbf{p}/m)^{\mu}{}_{\nu} = 
\left (
\begin{array}{cc}
\omega_m (\mathbf{p})/m & \mathbf{p}/m\\
\mathbf{p}/m & \delta_{ij} + {p_{i} p_{j} \over m (m+ 
\omega_m (\mathbf{p}))} \\ 
\end{array}
\right ).
\label{r.4}
\eeq
In (\ref{r.3}) and (\ref{r.4}) $\omega_m (\mathbf{p})= \sqrt{m^2 +
\mathbf{p}^2}$ is the energy of a particle of mass $m$ and momentum
$\mathbf{p}$.  These are related to the rapidity by 
$\cosh (\rho) = \omega_m (\mathbf{p})/m$ and 
$\sinh (\rho) = \vert \mathbf{p} \vert /m$.

The energy factors make (\ref{r.3}) unitary if the states, $\vert (m,j)
\mathbf{p},\mu \rangle$,  are given a delta-function normalization,
\beq
\langle (m,j)\mathbf{p}',\mu'
\vert (m,j)\mathbf{p},\mu \rangle
= \delta_{\mu'\mu} \delta (\mathbf{p}'-\mathbf{p}).
\label{r.5}
\eeq
From (\ref{r.4}) it follows that
\beq
p=  B (\mathbf{p}/m) (m,0,0,0) = (\omega_m (\mathbf{p}),\mathbf{p}) .
\eeq

A general Poincar\'e transformation, $U(\Lambda ,a)$,  
on a single-particle state, $\vert
(m,j)\mathbf{p},\mu \rangle$, can be decomposed into a product of the
three elementary unitary transformations (\ref{r.1}),(\ref{r.2}) and
(\ref{r.3}) using the group representation property
\beq
U(\Lambda ,a) = U(B (\pmb{\Lambda} p/m) \; 
T(B^{-1}(\pmb{\Lambda} p/m )a) \;
U (R_w(\Lambda , \mathbf{p}/m) ) \;
U (B^{-1}(\mathbf{p}/m)) 
\label{r.6}
\eeq
where 
\beq
R_w (\Lambda , \mathbf{p}/m):= 
B^{-1}(\pmb{\Lambda} p/m)\Lambda B(\mathbf{p}/m) 
\label{r.7}
\eeq
is a Wigner rotation.  

The decomposition (\ref{r.6}) is an inverse boost from a state with
momentum $\mathbf{p}$ to the rest state, followed by a rotation of the
rest state, followed by a translation of the rest state, and finishing
with a boost from the rest state to a state with the Lorentz
transformed momentum.

When the sequence of elementary transformations (\ref{r.6}) is applied to 
$\vert (m,j) \mathbf{p},\mu \rangle$
the result is  
\beq
U_{mj} (\Lambda,a ) \vert (m,j)\mathbf{p},\mu \rangle :=
\sum_{\nu=-j}^j  \vert (m,j)\pmb{\Lambda}{p},\nu \rangle
e^{i \Lambda p \cdot a} 
\sqrt{{\omega_m (\pmb{\Lambda} p) \over \omega_m (\mathbf{p})}}
D^j_{\nu\mu} \left[R_w(\Lambda ,\mathbf{p}/m)\right] 
\label{r.8}
\eeq
where the subscript $m,j$ indicates that this is a unitary
representation of the Poincar\'e group for a particle of mass $m$ and
spin $j$.  Eq.~(\ref{r.8}) defines mass $m$ spin $j$ unitary
irreducible representation of the Poincar\'e group.

It acts irreducibly on the Hilbert space ${\cal H}_{mj}$ spanned by
the single-particle states $\vert (m,j)\mathbf{p},\mu \rangle$.  The
irreducibility means that ${\cal H}_{mj}$ can be generated from any
fixed vector in ${\cal H}_{mj}$ by Poincar\'e transformations.

The construction used above to construct single-particle irreducible
representations will be used to construct $N$-particle irreducible
representations, which will be used in the construction of dynamical
irreducible representations.


\section{ $N$ non-interacting particles}

The Hilbert space for a system of $N$ non-interacting particles is the
$N$-fold tensor product of the single-particle Hilbert spaces
\beq
{\cal H} := \otimes_{i=1}^N {\cal H}_{m_i,j_i}  .
\label{r.9}
\eeq
For identical particles the physical Hilbert space is the projection
on the appropriately symmetrized or antisymmetrized subspace of
${\cal H}$.

The non-interacting (kinematic) unitary representation of the
Poincar\'e group on ${\cal H}$ is the tensor product of the
single-particle unitary representations of the Poincar\'e group
\beq
U_0 (\Lambda ,a) = \otimes_{i=1}^{N} U_{m_ij_i} (\Lambda ,a).
\label{r.10}
\eeq

A basis for the $N$-particle system is the direct product of the 
$N$ one-particle basis vectors
\beq
\vert \mathbf{p}_1, \mu_1, \cdots , \mathbf{p}_N, \mu_N \rangle :=
\prod_{l=1}^N  
\vert (m_l,j_l) \mathbf{p}_l,\mu_l \rangle ,
\label{r.11}
\eeq
where we have suppressed all of the single-particle mass and spin
quantum numbers on the left.

Following what we did in the single-particle case, we consider a basis
for the $N$-particle system in the rest frame of the $N$-particle
system.  We let $\mathbf{q}_i$ denote the momentum of the $i^{th}$
particle in the $N$-body rest frame.  The variables $\mathbf{q}_i$ are
constrained so
\beq
\sum \mathbf{q}_{i=1}^N = \mathbf{0} .
\label{r.12}
\eeq
We write the rest eigenstates as
\beq
\vert \mathbf{q}_1 , \mu_1, \cdots , \mathbf{q}_N, \mu_N \rangle 
\label{r.13}
\eeq
where it is understood that $\mathbf{q}_N = -\sum_{i \not= N}
\mathbf{q}_i$.  Following what we did for the single-particle states
we examine the rotational properties the rest eigenstates.

Using the transformation properties of the single-particle states
(\ref{r.8}) and the expression for $N$-particle Poincar\'e
transformations, in terms of the single-particle transformations
(\ref{r.10}), give the following transformation properties for the
$N$-particle rest eigenstates under rotations:
\[
U_0(R,0) \vert \mathbf{q}_1 , \mu_1, \cdots , \mathbf{q}_N, \mu_N \rangle =
\]
\beq
\sum_{\nu_1 \cdots \nu_N}  \vert R\mathbf{q}_1 , \nu_1, \cdots , R\mathbf{q}_N, \nu_N \rangle 
\prod_{l=1}^N D^{j_l}_{\nu_l \mu_l}\left(B^{-1}(R\mathbf{q}_l/m_l) R B (
\mathbf{q}_l/m_l)\right).
\label{r.14}
\eeq
The rotationless boosts have the distinguishing property that
\beq
B^{-1}(R\mathbf{q}/m) R B (
\mathbf{q}/m) = R
\label{r.15}
\eeq
for any $\mathbf{q}$.  This implies that ``the Wigner rotation of a
rotation is the rotation''.  It is a special property that is not
shared by other types of boosts.
 
As a consequence of this property (\ref{r.14}) becomes
\[
U_0(R,0) \vert \mathbf{q}_1 , \mu_1, \cdots , \mathbf{q}_N, \mu_N \rangle =
\]
\beq
\sum_{\nu_1 \cdots \nu_N} \vert R\mathbf{q}_1 , \nu_1, \cdots , R\mathbf{q}_N, \nu_N \rangle 
\prod_{l=1}^N D^{j_l}_{\nu_l \mu_l} (R) .
\label{r.16}
\eeq
This is exactly how a non-relativistic $N$-particle state transforms
under rotations.  It follows that all of the spins and orbital angular
momenta can be added with ordinary SU(2) Clebsch-Gordan coefficients
and spherical harmonics.  The primary difference with the
single-particle case is that there can be many orthogonal rotationally
invariant subspaces with the same $j$.  They are distinguished by
internal spins, orbital angular momenta and sub-energies.

The result is that the rest state can be decomposed into an orthogonal
direct sum of states with different total spin.  Since there are many
possible orders of coupling we denote these states by
\beq
\vert (M_0,j) \mathbf{0}, \mu ; \mathbf{d} \rangle ,
\label{r.17}
\eeq
where 
\beq
M_0 = \sum_{l=1}^N \sqrt{\mathbf{q}_l^2 +m_l^2} 
\label{r.18}
\eeq
is the invariant mass (rest energy) of this system and $\mathbf{d}$
are invariant degeneracy quantum numbers that distinguish different subspaces
with the same value of $j$.

For a two-body system with spin $j$ typical degeneracy parameters
would be $\mathbf{d}= \{l,s\}$.  For a three-particle system we could
have $l_{ij},s_{ij},j_{ij},k_{ij}$ for the $ij$ pair, where $k_{ij}$ is the
magnitude of the rest momentum of the $ij$ pair, and
$L_{ij,k},S_{ij,k}$ representing the orbital and spin quantum numbers
associated with the pair and third particle.  In this case
$\mathbf{d}= \{ l_{ij},s_{ij},j_{ij},k_{ij}, L_{ij,k},S_{ij,k}\}$.

The choice of degeneracy parameters is normally made for convenience;
for example the three-body choice above would be useful for
constructing matrix elements of an interaction between particles $i$
and $j$.  The important observation is that they are all rotationally
invariant quantum numbers. In general $\mathbf{d}$ includes both
discrete quantum numbers like $l_{ij},s_{ij},L_{ij,k},S_{ij,k}$ and
continuous ones like $k_{ij}$.  Different choices of $\mathbf{d}$ are
related by unitary transformations whose coefficients are Racah
coefficients for the Poincar\'e group. 

The result of coupling the spins means that in this basis (\ref{r.16})
has the same form as (\ref{r.1}):
\beq
U_0(R,0) \vert \mathbf{0}, M_0, j, \mu ;\mathbf{d} \rangle  =
\sum_{\nu=-j}^j \vert \mathbf{0}, M_0, j, \nu ;\mathbf{d} \rangle  
D^j_{\nu \mu}(R).
\label{r.19}
\eeq
The differences are the presence of the invariant degeneracy 
parameters $\mathbf{d}$ and the fact that the invariant mass
$M_0$ has a continuous spectrum that runs from the sum of the 
individual masses to infinity.  

The states $(\ref{r.17})$ are rest states.  We can define states
with a non-zero total momentum and the same spin by analogy with 
(\ref{r.3})
\beq
\vert (M_0,j)\mathbf{P},\mu,\mathbf{d} \rangle := 
U_0(B (\mathbf{P}/M_0)) \vert (M_0,j)\mathbf{0},\mu; \mathbf{d}  \rangle
\sqrt{{M_0 \over \omega_{M_0} (\mathbf{P})}}.
\label{r.20}
\eeq
The difference between this equation and (\ref{r.3}) is that
(\ref{r.3}) was used to define the unitary representation of the
rotationless boost, while in this case the representation of the
rotationless boost is given by (\ref{r.10}) so (\ref{r.20}) defines
the momentum-spin eigenstate.  This definition implies a 
delta-function normalization in $\mathbf{P}$.  It redefines the magnetic
quantum numbers so they agree with the single-particle magnetic
quantum numbers when boosted to the rest frame of the $N$-particle
system with a rotationless boost.

Unitarity gives the normalization
\beq
\langle  (M_0',j')\mathbf{P}',\mu';\mathbf{d}' 
\vert (M_0,j)\mathbf{P},\mu;\mathbf{d} \rangle
=
\delta (M_0'-M_0)\; \delta (\mathbf{P}-\mathbf{P}') 
\; \delta_{j'j} \delta_{\mu' \mu} \delta_{\mathbf{d}':\mathbf{d}}
\label{r.21}
\eeq
where $\delta_{\mathbf{d}':\mathbf{d}}$ is a product of Dirac delta 
functions in the continuous degeneracy quantum numbers and Kronecker 
delta functions in the discrete degeneracy quantum numbers.

It is not hard to show that (\ref{r.20}) is an eigenstate of the total
momentum.  The same steps used in (\ref{r.8}) lead to the following
unitary representation of the Poincar\'e group for the non-interacting
system,
\beq 
U_0(\Lambda ,a) \vert (M_0,j)\mathbf{P},\mu; \mathbf{d}  \rangle =
\sum_{\nu=-j}^j  \vert (M_0,j)\pmb{\Lambda}{P},\nu;\mathbf{d} \rangle
e^{i \Lambda P \cdot a} 
\sqrt{{\omega_{M_0} (\pmb{\Lambda} P) \over \omega_{M_0} (\mathbf{P})}}
D^j_{\nu\mu} \left[ R_w(\Lambda ,\mathbf{P}/M_0) \right] .
\label{r.22}
\eeq
In constructing this basis we have decomposed products of
irreducible representations of the Poincar\'e group into orthogonal
direct integrals of irreducible representations.  The coefficients of
this transformation are the Clebsch-Gordan coefficients for the
Poincar\'e group.
 
It is instructive to see the form of these coefficients in a specific
example.  We consider the case of coupling two particles.  In that
case the two-body rest state (\ref{r.13}) is
\beq
\vert \mathbf{q}_1, \mu_1 ,-\mathbf{q}_1 , \mu_2 \rangle 
\label{r.23}
\eeq
where we have used the constraint $\mathbf{q}_1 + \mathbf{q}_2=0$.
The decomposition of (\ref{r.17}) into irreducible representations of the
rotation group is
\beq
\vert (M_0, j ) \mathbf{0}, \mu ; l,s \rangle :=
\sum_{\mu_1 \mu_2 m \mu_s} \int d\hat{\mathbf{q}}_1
\vert \mathbf{q}_1, \mu_1 ,-\mathbf{q}_1 , \mu_2 \rangle \;
Y_{lm}( \hat{\mathbf{q}}_1)
\langle j_1, \mu_1, j_2, \mu_2, \vert s, \mu_s \rangle
\langle s, \mu_s, l, m \vert j, \mu \rangle ,
\label{r.24}
\eeq
where
\beq
M_0 = \sqrt{\mathbf{q}_1^2 +m_1^2 } + \sqrt{\mathbf{q}_1^2 + m_2^2}.
\label{r.25}
\eeq
The $Y_{lm}( \hat{\mathbf{q}}_1)$ are spherical harmonics and $\langle j_1,
\mu_1 , j_2, \mu_2 \vert j_3, \mu_3 \rangle$ are $SU(2)$ Clebsch-Gordan
coefficients.  Applying a rotationless boost to both side of equation
(\ref{r.24}), using (\ref{r.8}) and (\ref{r.10}) on the right and
(\ref{r.20}) on the left gives
\begin{eqnarray}
\lefteqn{\vert (M_0, j )\mathbf{P}, \mu ; l,s \rangle :=}&&  \cr
& & \sum_{\nu_1,\nu_2,\mu_2,\mu_2,\mu_s,m} \int \hat{\mathbf{q}}_1
\vert \mathbf{p}_1, \nu_1 ,\mathbf{p}_2 , \nu_2 \rangle
\sqrt{{\omega_{m_1} (\mathbf{p}_1)  \over
\omega_{m_1} (\mathbf{q}_1)}}
\sqrt{{\omega_{m_2} (\mathbf{p}_2)  \over
\omega_{m_2} (\mathbf{q}_1)}} \times \; \cr
& & 
D^{j_1}_{\nu_1 \mu_1}[B^{-1}(\mathbf{p}_1/m_1) B(\mathbf{P}/M_0)
B(\mathbf{q}_1/m_1)] \times \cr
& & D^{j_2}_{\nu_2 \mu_2} \left[B^{-1}(\mathbf{p}_2/m_2) B(\mathbf{P}/M_0)
B(-\mathbf{q}_1/m_2)\right] \times \; \cr
& & Y_{lm}( \hat{\mathbf{q}}_1) \;
 \langle j_1, \mu_1, j_2, \mu_2, \vert s, \mu_s \rangle
\langle s, \mu_s, l, m \vert j, \mu \rangle \;
\sqrt{{M_0 \over \omega_{M_0} (\mathbf{P})}}
\label{r.26}
\end{eqnarray}
where $\mathbf{q}_i$ and $\mathbf{p}_i$ are related by
\beq
q_i = B^{-1}(\mathbf{P}/M_0) p_i
\label{r.27}
\eeq
which can be expressed in terms of the three-vector components using
(\ref{r.4}) as
\beq
\mathbf{q}_i = \mathbf{p}_i + {\mathbf{P}\over M_0}
\left ( {\mathbf{P} \cdot \mathbf{p}_i \over
M_0 + \omega_{M_0}(\mathbf{P})} - \omega_{m_i}(\mathbf{p}_i)\right ). 
\label{r.28}
\eeq
The sums in (\ref{r.26}) are over the magnetic quantum numbers
$\nu_1,\nu_2,\mu_1,\mu_2,\vert j_1-j_2\vert  \leq s \leq \vert j_1+j_2\vert,
\vert j-s\vert \leq l \leq \vert j+s \vert$ and the orbital 
magnetic quantum number $m$.

The Poincar\'e group Clebsch-Gordan coefficients are the coefficients
of the unitary transformation (\ref{r.26}).

Returning to the $N$-particle case, note that the boost acts on the 
state in equation (\ref{r.20}) while the transformation between
$\{M_0,j,\mu,\mathbf{d}\}$ and $\{ \mathbf{q}_1, \mu_1, \cdots ,
\mathbf{q}_N, \mu_N \}$ acts on the quantum numbers.  
The result of transforming
the variables on right side of (\ref{r.20}) leads to
\beq 
\vert \mathbf{P}; \mathbf{q}_1, \mu_1, \cdots ,\mathbf{q}_N,\mu_N  
\rangle := 
U(B(\mathbf{P}/M_0) 
\vert \mathbf{q}_1, \mu_1, \cdots ,\mathbf{q}_N,\mu_N  
\rangle 
\sqrt{{M_0 \over \omega_{M_0}(\mathbf{P})}}.
\label{r.29}
\eeq
The relation of these states to the original single-particle states
follows from (\ref{r.8}),(\ref{r.10}) and (\ref{r.29}):
\begin{eqnarray}
\lefteqn{\vert \mathbf{P}; \mathbf{q}_1, \mu_1, \cdots ,\mathbf{q}_N,\mu_N  
\rangle := } \cr
& & \sum_{\nu_1 \cdots \nu_N}  
\vert \mathbf{p}_1, \nu_1, \cdots ,\mathbf{p}_N, \nu_N \rangle
\sqrt{{M_0 \over \omega_{M_0}(\mathbf{P})}}\;
\prod_{k=1}^N D^{j_k}_{\nu_k \mu_k}[B^{-1}(\mathbf{p}_k/m_k) B(\mathbf{P}/M_0)
B(\mathbf{q}_k/m_k)] \times \cr
& & \sqrt{{\omega_{m_k} (\mathbf{p}_k)  \over \omega_{m_k} (\mathbf{q}_k)}} 
\label{r.30}
\end{eqnarray}
where the $p_i$ are related to the $q_i$ by (\ref{r.28}).  There is a
corresponding relation between the spins implied by (\ref{r.30}).  We
refer to the spins, $\mu_1 \cdots \mu_N$, on the left side of (\ref{r.30}) 
as constituent
single-particle spins and the spins, $\nu_1 \cdots \nu_N$,  
on the right as single-particle
spins.  The corresponding spin operators are related by Wigner
rotations
\beq
(0,\mathbf{j}_{ic} )  
=  B^{-1} (\mathbf{q}_i /m_i) 
B^{-1} (\mathbf{P} /M_0) B (\mathbf{p}_i/ m_i) 
(0,\mathbf{j}_{i} ).
\label{r.31}
\eeq
These spins become identical in the $N$-particle rest frame.  The
constituent spins have the advantage that they remain unchanged under
boosts from the $N$-body rest frame and they all experience the same
Wigner rotation under general Lorentz transformations.  The advantage
of using a basis with constituent spins is that they can be added like
non-relativistic spins.


\section {$N$ interacting particles}

In this section we construct a dynamical unitary representation of the
Poincar\'e group.  We use two equivalent constructions - one is
designed to provide an explicit representation of the dynamical
unitary representation of the Poincar\'e group while the other is more
appropriate for $N$-particle applications.  We start with the
construction of the explicit representation of the dynamical unitary
representation of the Poincar\'e group.

The simplest way to construct a relativistic $N$-particle dynamics is
to start with the non-interacting $N$-particle irreducible basis
(\ref{r.22}) constructed in the previous section
\beq
\vert (M_0 ,j ) \mathbf{P}, \mu, \mathbf{d} \rangle .
\label{r.32}
\eeq
In order to construct an interacting unitary irreducible
representation of the Poincar\'e group we add an interaction $V$ to
$M_0$ that commutes with the non-interacting spin,
$\mathbf{j}$,
\beq
M= M_0 +V.
\label{r.33}
\eeq 
We also assume that $V$ is translationally invariant and is
independent of the total momentum.

A general interaction of this form has matrix elements in the $N$ 
free-particle irreducible basis (\ref{r.22}) of the form
\[
\langle (M_0',j') \mathbf{P}',\mu' ; \mathbf{d}'  \vert V
\vert  (M_0,j) \mathbf{P},\mu ; \mathbf{d}  \rangle =
\]
\beq
\delta (\mathbf{P}'-\mathbf{P}) 
\delta_{j'j} \delta_{\mu'\mu} 
\langle M_0', \mathbf{d}'  \Vert V^{j} \Vert 
M_0, \mathbf{d} \rangle .
\label{r.34}
\eeq
For two particles $M_0 = \sqrt{\mathbf{q}^2 + m_1^2} +
\sqrt{\mathbf{q}^2 + m_2^2}$ where $\mathbf{q}$ is the rest-frame
momentum of particle 1 and the degeneracy parameters, $l^2$ and $s^2$,
are orbital and spin angular momenta, so with a suitable change of
variables (\ref{r.34}) looks like a standard two-body interaction in a
partial-wave representation.

Simultaneous eigenstates of $M$, $\mathbf{P}$, $\mathbf{j}^2$ and
$\hat{\mathbf{z}} \cdot \mathbf{j}$ can be constructed by
diagonalizing $M$ in the basis of eigenstates of $M_0$, $\mathbf{P}$,
$\mathbf{j}^2$ and $\hat{\mathbf{z}} \cdot \mathbf{j}$.

The symmetry properties of the interaction (\ref{r.34}) 
imply that eigenfunctions have the form 
\beq
\langle (M_0,j) \mathbf{P},\mu ; \mathbf{d}  \vert 
(\lambda ,j') ,\mathbf{P}',
\mu' \rangle 
=\delta (\mathbf{P}-\mathbf{P}') \delta_{jj'}
\delta_{\mu \mu'} 
\psi_{\lambda ,j} (\mathbf{d},M_0) ,
\label{r.35}
\eeq
where the wave functions, $\psi_{\lambda ,j} (\mathbf{d},M_0)$, are
solutions to the relativistic mass eigenvalue problem
\beq
(\lambda -M_0  )\psi_{\lambda ,j} (\mathbf{d},M_0)  =
\sum' \int dM_0' d\mathbf{d}' 
\langle M_0,\mathbf{d} \vert V^{j} \vert M_0',\mathbf{d}' \rangle  
\psi_{\lambda ,j} (\mathbf{d}',M_0') 
\label{r.36}
\eeq
and $\lambda$ is the mass eigenvalue.  Here the sum is over the discrete
degeneracy quantum numbers, the integrals are over the 
continuous degeneracy quantum numbers and the 
spectrum of the 
invariant mass operator $M_0$.
This equation replaces the
many-body Schr\"odinger equation for the center-of-mass Hamiltonian in
non-relativistic quantum mechanics. The eigenstates 
\beq
\vert (\lambda ,j) ,\mathbf{P},
\mu \rangle
\label{r.37}
\eeq 
transform like (\ref{r.22}) with the mass eigenvalue 
$\lambda$ replacing $M_0$ in (\ref{r.22}):
\[
U(\Lambda,a ) \vert (\lambda,j)\mathbf{P},\mu \rangle =
\]
\beq
\sum_{\nu=-{j}}^{j}  \vert (\lambda,j)\pmb{\Lambda}{P},\nu  
\rangle 
e^{i \Lambda P \cdot a} 
\sqrt{{\omega_{\lambda} (\pmb{\Lambda} P_0) \over \omega_{\lambda} 
(\mathbf{P}_0)}}
D^{j}_{\nu\mu} [R_w (\Lambda, \mathbf{P}/\lambda )], 
\label{r.38}
\eeq
where in this case the Wigner rotation depends on the mass 
eigenvalue, $\lambda$, 
\beq
R_w (\Lambda, \mathbf{P}/\lambda) = B^{-1} (\pmb{\Lambda}P/\lambda) 
\Lambda B(\mathbf{P}/ \lambda)  \qquad P^0=\sqrt{\lambda^2 + \mathbf{P}^2} .
\label{r.39}
\eeq
In these expressions $P^{\mu}$ is the four-momentum of the interacting
system, which has different mass and energy eigenvalues than
the non-interacting system.
A complete set of irreducible eigenstates will have multiple copies of
states with the same mass and spin that are distinguish by invariant
degeneracy quantum numbers.  Since the eigenstates
(\ref{r.37}) are complete, (\ref{r.38}) defines the
dynamical unitary representation of the Poincar\'e group on ${\cal
  H}$.

This shows that the construction of the dynamical representation of
the Poincar\'e group can be reduced to solving the mass eigenvalue
problem (\ref{r.36}).  This is analogous to constructing the unitary
time evolution operator by diagonalizing the center of mass
Hamiltonian in non-relativistic quantum mechanics.

This construction was first performed by Bakamjian and Thomas
\cite{bakamjian:1951} for the two-particle system.  For systems of
more than two particles this construction fails to satisfy cluster
properties \cite{fcwp:1982,bkwp:2012}, which means that
\beq
U(\Lambda,a) \not\to U_I(\Lambda,a) \otimes U_{II}(\Lambda,a)  
\label{r.40}
\eeq
on states corresponding to asymptotically separated subsystems, I and
II. 

This deficiency can be systematically corrected \cite{wkwp}: the
corrections appear in the form of additional many-body interactions
that are functions of the input interactions.  The interactions that
restore cluster properties fall-off like powers of 
$(V/m)^{N-1}$~\cite{fcwp:1982}, where $V$ is the two-body interaction. They appear to be
small in nuclear physics applications~\cite{bkwp:2012}. Thus in the following 
these corrections will be ignored.

While the $N$ free-particle irreducible basis is the most convenient for
illustrating the construction of a dynamical unitary representation of
the Poincar\'e group, like a partial-wave basis, it is not an ideal
basis for many-body problems.  In addition, for relativistic problems
partial-wave expansions can lead to numerical challenges~\cite{Lin:2008sy}.

Note that the rest states (\ref{r.13}) and (\ref{r.17}) only differ by
an ordinary partial-wave expansion constructed out of linear
combinations of these states with different arguments, while states
with arbitrary momentum are constructed by applying a unitary boost to
these linear combinations, that leaves all of the quantum numbers
unchanged except the total momentum.

This implies that the $N$-body basis 
\beq
\vert \mathbf{P}; \mathbf{q}_1,
\mu_1, \cdots \mathbf{q}_N, \mu_N \rangle 
:= 
U_0(B (\mathbf{P}/M_0))\vert \mathbf{q}_1, 
\mu_1, \cdots \mathbf{q}_n, \mu_n \rangle 
\sqrt{{M_0 \over \omega_{M_0} (\mathbf{P})}}
\label{r.41}\eeq
is related to (\ref{r.20}) by SU(2) Clebsch-Gordan coefficients and
spherical harmonics.

In the basis (\ref{r.41}) the interaction can be expressed as
\begin{eqnarray}
\lefteqn{\langle \mathbf{P},  \mathbf{q}_1 , \mu_1, \cdots , \mathbf{q}_N, \mu_N 
\vert V \vert \mathbf{P}', \mathbf{q}'_1 , 
\mu'_1, \cdots , \mathbf{q}'_N, \mu'_N 
\rangle =} \cr
& &\delta (\mathbf{P}-\mathbf{P}') 
\langle \mathbf{q}_1 , \mu_1, \cdots ,  \mathbf{q}_N, \mu_N 
\Vert V \Vert  \mathbf{q}'_1 , \mu'_1, \cdots , \mathbf{q}'_N, \mu'_N 
\rangle , 
\label{r.42}
\end{eqnarray}
where rotational invariance means that the reduced kernel satisfies
\begin{eqnarray}
\lefteqn{\langle  \mathbf{q}_1 , \mu_1, \cdots, \mathbf{q}_N, \mu_N 
\Vert V \Vert \mathbf{q}'_1 , \mu'_1, \cdots , \mathbf{q}'_N, \mu'_N 
\rangle =} \cr
& & 
\sum_{\nu_1 \cdots \nu_N,\nu_1' \cdots \nu_N'} \prod_{i=1}^N 
D^{j_i}_{\mu_i \nu_i} [R^{-1}] 
\langle  R\mathbf{q}_1 , \nu_1, \cdots , R\mathbf{q}_N, \nu_N 
\Vert V \Vert R\mathbf{q}'_1 , \nu'_1, \cdots , R\mathbf{q}'_N, \nu'_N 
\rangle \times \cr
& & 
\prod_{l=1}^N D^{j_l}_{\nu_l' \mu_l'} [R] 
\label{r.43}
\end{eqnarray}
for any rotation $R$.  The only other requirements on $V$ are
$V=V^{\dagger}$ and $M_0 + V > 0$.

In this representation a general interaction is a sum of $2,3,4
\cdots$-body interactions.
The mass eigenfunctions (\ref{r.35}) have the form  
\beq
\langle \mathbf{P},  \mathbf{q}_1 , \mu_1, \cdots , \mathbf{q}_N, \mu_N
\vert  (\lambda , j ) \mathbf{P}', \mu' \rangle =
 \delta (\mathbf{P}-\mathbf{P}')  
\langle \mathbf{q}_1 , \mu_1, \cdots , \mathbf{q}_N, \mu_N
\vert  (\lambda,j) \mu' \rangle , 
\label{r.44}
\eeq
and the mass eigenvalue problem (\ref{r.36}) has the form
\begin{eqnarray}
\lefteqn{\left(\lambda - \sum_i \sqrt{\mathbf{q}_i^2 +m_i^2} \right)
\langle \mathbf{q}_1 , \mu_1, \cdots ,\mathbf{q}_N, \mu_N
\vert (\lambda,j) \mu \rangle = } \cr
& & 
\sum_{\mu_1' \cdots \mu_N'}  \int
\langle \mathbf{q}_1 , \mu_1, \cdots ,\mathbf{q}_N, \mu_N 
\Vert V \Vert \mathbf{q}'_1 , \mu'_1, \cdots , \mathbf{q}'_N, \mu'_N 
\rangle \;
d\mathbf{q}'_1 \cdots \mathbf{q}'_N \; \times \cr 
& & \delta \left(\sum_{i=1}^N \mathbf{q}'_i\right) 
\langle \mathbf{q}_1' , \mu_1', \cdots , \mathbf{q}_N', \mu_N'
\vert  (\lambda,j) \mu \rangle .
\label{r.45}
\end{eqnarray}
The relativistic transformation properties can be easily determined
once $M$ is diagonalized.  These eigenstates transform like mass
$\lambda$ spin $j$ irreducible representations (\ref{r.38}).


\section{Reaction theory models}

For most nuclear systems a direct solution of the quantum mechanical
scattering problem is not feasible.  Approximations that are dominated
by a more limited number of degrees of freedom are often amenable to a
numerical solution.  Success depends on identifying the most important
degrees of freedom.  In addition the effective interactions need to be
modeled.  Nevertheless it is useful to have a formalism where this
is the first step in a well-defined systematic approximation to the
exact solution.

To formulate a relativistic reaction model the steps are (1) start
with an exact relativistic quantum mechanical model, (2) identify the
most important degrees of freedom and then (3) construct an
approximate relativistic quantum mechanical model with those degrees
of freedom.  To do this we project the exact mass operator on a
relativistically invariant coupled-channel subspace of the full
Hilbert space that allows scattering in all of the chosen important
reaction channels.  The relativistic invariance is preserved by
choosing the projection to have the same symmetries as the
interaction.  The relation to the full theory provides a means to
systematically include additional degrees of freedom.

The starting point is a relativistic mass operator (\ref{r.33}) (or
rest energy operator) which in the basis (\ref{r.29}) has the form
\beq
M = \sum_{i=1}^N \sqrt{\mathbf{q}_i^2 + m_i^2} +
\sum_{i< j}^N  V_{ij} + \sum_{i<j<k}^N V_{ijk} + \cdots
\label{r.46}
\eeq
where the sum of the $\mathbf{q}_i$ add to zero and the interactions
are rotationally invariant operators that depend on the $\mathbf{q}_i$
and the constituent spins (\ref{r.30}-\ref{r.31}).

For any partition $a$ of the $N$-particle system into disjoint
subsystems we construct the partition mass operator $M_a$ by
eliminating interactions that involve particles in different clusters
of the partition $a$.  We also define the residual interactions
\beq
V^a := M - M_a .
\label{r.47}
\eeq
The operator $M_a$ is a sum of operators $M_{a_k}$ for each disjoint
non-empty cluster, $a_k$ of $a$:
\beq
M_a = \sum_k M_{a_k} 
\label{r.48}
\eeq
given by 
\beq
M_{a_k} = \sum_{i\in a_k} \sqrt{\mathbf{q}_i^2 + m_i^2} +
\sum_{i<j\in a_k} V_{ij} + \sum_{i<j<l\in a_k} V_{ijl} +\cdots .
\label{r.49}
\eeq
In these expressions the $\mathbf{q}_i$ are not constrained in the
various subsystems, however the total momentum of the subsystems is
constrained to add up to zero only in the $N$-body system.  This means
that the operators $M_{a_k}$ represent the energy of the moving
clusters in the $N$-body rest frame.
 
$M_{a_k}$ has the same form as (\ref{r.46}) except the sum is only over 
the particles in the $k^{th}$ cluster of $a$.  The natural variables
for the for solving the subsystem problem are the subsystem
constituent spins and the subsystem rest momenta $\mathbf{k}_i$.   
These are related to the system constituent spins and rest
momenta by a relation like (\ref{r.30})
\begin{eqnarray}
\lefteqn{\vert \mathbf{q}_{ak1}, \mu_{ak1}, \cdots ,\mathbf{q}_{akl}, \mu_{akl} \rangle
=
\sum_{\nu_1 \cdots \nu_l}  
\vert \mathbf{q}_{a_k} , \mathbf{k}_1, \nu_1, \cdots ,
\mathbf{k}_l, \nu_l \rangle \times} \cr
& & 
\sqrt{{\sum_{r\in a_k}\omega_{m_{akr}} (\mathbf{q}_r)\over 
\sum_{s\in a_k} \omega_{m_{aks}} (\mathbf{k}_s)}}
\prod_{i=1}^l 
D_{\nu_i\mu_{aki}} [(B^{-1}(\mathbf{k}_i/m_{aki}) B^{-1}(\mathbf{q}_a/ M_{0a})
B(\mathbf{q}_{aki} /m_{aki})] 
\sqrt{{\omega_{m_{aki}} (\mathbf{k}_i)\over \omega_{m_{aki}} (\mathbf{q}_i)}}
\label{r.50}
\end{eqnarray}
where 
\beq
\mathbf{q}_{a_k}=
\sum_{i\in a_k} \mathbf{q}_i,
\qquad
k_i := B^{-1} (\mathbf{q}_{a_k} /M_{ak0})q_i,
\qquad
\sum_{i\in a_k} \mathbf{k}_i=0
\label{r.51}
\eeq
and $M_{ak0}$ is the invariant mass of the non-interacting
subsystem.   These wave functions have the same form as an 
$N$-body bound state in the basis (\ref{r.11}), except the 
$\mathbf{p}_i$ are replaced by the corresponding $\mathbf{q}_i$ 
and the single particle spins are replaced by the constituent
single-particle spins.  
When these are embedded in the full
system the sum of the cluster momenta, $\sum_k \mathbf{q}_{a_k} = 0 $,
are constrained to add to zero.
 
Each of the cluster mass operators, $M_{a_k}$, will have simultaneous
eigenstates of $\mathbf{q}_{a_k}$ and subsystem mass $\lambda_{a_k}$.
For the purpose of reaction theory we are interested only in the case
that $\lambda_{a_k}$ are point-spectrum eigenvalues corresponding to
bound clusters.  In the $n_k$-free particle basis variables these
subsystem mass eigenstates have the form
\begin{eqnarray}
\lefteqn{\langle \mathbf{q}_{k1}, \mu_{k1}, \cdots ,\mathbf{q}_{kn_k}, \mu_{kn_k} 
\vert  (\lambda_{a_k}, j_{a_k} ) \mathbf{q}_{a_k}, \mu_{a_k} \rangle = }\cr
& & 
\delta \left(\sum_{i=1}^{n_k} \mathbf{q}_{ki} - \mathbf{q}_{a_k}\right) 
\langle \mathbf{q}_{k1}, \mu_{k1}, \cdots ,\mathbf{q}_{kn_k}, \mu_{kn_k} 
\vert \lambda_{a_k}, j_{a_k} ;\mathbf{q}_{a_k}, \mu_{a_k} \rangle 
\label{r.52}
\end{eqnarray}
In this expression (\ref{r.50}) is used to relate the subsystem
variables to the variables of the basis (\ref{r.30}).

Channel projection operators can be defined in terms of products of
these eigenstates:
\beq
\Pi_{\alpha} = 
\prod_j
\int \sum_{\mu_{a_j}}
\vert (\lambda_{a_j}, j_{a_j}) \mathbf{q}_{a_j}, \mu_{a_j} \rangle 
d\mathbf{q}_{a_j} \; \delta \left(\sum_l \mathbf{q}_{a_l} \right)
\langle  (\lambda_{a_j}, j_{a_j}) \mathbf{q}_{a_j}, \mu_{a_j} \vert
\; \vert \mathbf{P} \rangle d \mathbf{P} \langle \mathbf{P} \vert 
\label{r.53}
\eeq
where the product is over all subsystems $a_j$ in a given partition
$a$ of the $N$-particle system and the additional index $\alpha$
indicates both the partition into bound subsystems as well as the
specific collection of bound states associated with each subsystem.

These channel projectors are used to build a projection on the
model space.

To construct a relativistic reaction theory we project the mass
operator on a subspace of the Hilbert space using projection operators
$\Pi_\alpha$ that commute with $\mathbf{P}$, are independent of $\mathbf{P}$
and commute with $\mathbf{j}$,
\beq
M_\pi := \Pi M \Pi .
\label{r.54}
\eeq

Simultaneous eigenstates of the projected mass operator, $M_\pi$, and
$\mathbf{P}$, $\mathbf{j}$ and $\mathbf{\hat{z}} \cdot \mathbf{j}$
transform like (\ref{r.38}) with respect to the Poincar\'e group.
This defines the relativistic model in terms of a unitary
representation of the Poincar\'e group on the model space.

The projection operator is a relativistic version of the projection
operators that appear in coupled-channel approximations.  It is
constructed from elementary projection operators that project on
subspaces generated by disjoint subsystems, where particles in the
same subsystem are bound and the bound subsystems are free to move
like free particles.  The subsystem bound states are solutions to
relativistic eigenvalue problems of the form (\ref{r.36}) with
$\lambda$ being a point-spectrum eigenvalue of the subsystem mass
operator.

The first step in making a reaction model is usually to construct the
projection operator $\Pi=\Pi_{\cal C}$ corresponding to a chosen set
of dominant reaction channels, ${\cal C}$.  Typically, if $\alpha \in
{\cal C}$ then it is normal to also include all channels generated
from the channel $\alpha$ by exchange of identical particles.

The sum $\Sigma_{\cal C}$ of the channel projectors over the subset
${\cal C}$ of scattering channels is the positive self-adjoint
operator
\beq
\Sigma_{\cal C} :=\sum_{\alpha \in {\cal C}} \Pi_\alpha .
\label{r.55}
\eeq

The main ideas that underly the formalism below were developed in a
series of papers by Chandler and Gibson \cite{cg}.
Let $\Sigma_{\cal C}^{\#}$ be the Moore-Penrose generalized inverse of
$\Sigma_{\cal C}$.  It is the unique solution to the Penrose equations
\cite{ben_israel}:
\begin{eqnarray}
(\Sigma_{\cal C}^{\#} \Sigma_{\cal C})& =& (\Sigma_{\cal C}^{\#} \Sigma_{\cal
C})^{\dagger}  \nonumber \\
\label{r.56}
(\Sigma_{\cal C} \Sigma_{\cal C}^{\#})& =& (\Sigma_{\cal C} 
\Sigma_{\cal C}^{\#})^{\dagger} 
\label{r.57} \nonumber \\
\Sigma_{\cal C} \Sigma_{\cal C}^{\#} \Sigma_{\cal C}& = &\Sigma_{\cal C}
\label{r.58} \nonumber \\
\Sigma^{\#}_{\cal C} \Sigma_{\cal C} \Sigma_{\cal C}^{\#}& =& \Sigma_{\cal C}^{\#}.
\label{r.59}
\end{eqnarray}
Because $\Sigma_{\cal C}=\Sigma_{\cal C}^{\dagger}$ it follows that 
\beq
[\Sigma_{\cal C},\Sigma_{\cal C}^{\#}]=0 
\label{r.60}
\eeq
and 
\beq
\Pi_{\cal C} =  \Sigma_{\cal C} \Sigma_{\cal C}^{\#} 
= \Sigma_{\cal C}^{\#}\Sigma_{\cal C}
\label{r.61}
\eeq
is an orthogonal projector on the subspace of the Hilbert space satisfying
\beq
\Pi_\alpha \Pi_{\cal C} = \Pi_{\cal C} \Pi_\alpha = \Pi_{\alpha}.
\label{r.62}
\eeq
In addition,  if $\vert x \rangle$ is any vector orthogonal to the range of 
$\Pi_\alpha$, 
\beq
\Pi_\alpha \vert x \rangle  = 0 
\label{r.63}
\eeq
for all $\alpha \in {\cal C}$ then 
\beq
\Pi_{\cal C} \vert x \rangle  = 0 .
\label{r.64}
\eeq
The results above follow because the range of $\Sigma_{\cal C}$
contains the range of $\Pi_\alpha$.  To show this assume that $\vert x
\rangle$ is in the range of $\Pi_\alpha$ for some $\alpha \in {\cal C}$ 
but $\vert x \rangle$ is not
in the range of $\Sigma_{\cal C}$.  It follows that
\beq
0 = \langle x \vert \Sigma_{\cal C} \vert x \rangle  = 
\langle x  \vert x \rangle   
+\sum_{\alpha'\not=\alpha}   
\langle x \vert \Pi_{\alpha'}   \vert x \rangle
\geq  \Vert \vert x \rangle \Vert^2 >0 
\label{r.65}
\eeq
which is a contradiction.
This shows that $\Pi_{\cal C}$ is an orthogonal projector on the smallest
subspace containing all of the channel subspaces in ${\cal C}$.
Some methods to compute the Moore-Penrose generalized inverse are 
discussed in
Appendix~\ref{appendixA}


\section{Relativistic Scattering Theory}

This section derives the symmetrized $S$ matrix for a relativistic
mass operator projected on a subspace that allows scattering in a
limited number of channels.  Rather than working on the model Hilbert
space defined on by range of $\Pi_{\cal C}$, it is useful to work on
the asymptotic channel spaces.  This has the advantage that the
dynamical equations only involve transition matrix elements projected
on the appropriate asymptotic states and interactions smeared with
subsystem bound-state wave functions.  This leads to a slightly
different type of coupled integral equations, where only the projected
part to the transition operators appear in the equations.  This is an
important simplification for reaction models because the projection of
the transition operator on unphysical subspaces do not appear in the
equations.

The relativistic reaction theory is the approximate theory defined by
replacing the exact mass operator by the projected mass operator
\beq
M \to M_{\Pi} = \Pi_{\cal C} M \Pi_{\cal C} . 
\label{r.66}
\eeq
The set of retained channels ${\cal C}$ is assumed to be invariant
with respect to permutations.  For this choice $M_{\Pi}$ commutes with
the symmetrizer (antisymmetrizer)  $A$,
\beq
[M_{\Pi},A]=0.
\label{r.67}
\eeq
In order to formulate scattering asymptotic conditions for each 
channel $\alpha\in {\cal C}$ there is a natural asymptotic Hilbert space
defined as the tensor product of irreducible representation spaces
associated with the mass and spin of each bound cluster in the 
channel $\alpha$;
\beq
{\cal H}_{\alpha} := \otimes_{j \in \alpha} {\cal H}_{\lambda_j j_j}.  
\label{r.68}
\eeq
The product of the irreducible state vectors in the channel $\alpha$
defines a mapping from ${\cal H}_{\alpha}$ to the model Hilbert space
${\cal H}_{\Pi}$ (the
range of $\Pi_{\cal C}$):
\beq
\Phi_{\alpha}:{\cal H}_{\alpha} \to {\cal H}_{\Pi}
\label{r.69}
\eeq
given by 
\beq
\Phi_{\alpha} \vert f_\alpha \rangle  := 
\int 
\prod_{j=1}^{m}
\sum_{\mu_{a_j}}
\vert (\lambda_{a_j}, j_{a_j}) \mathbf{q}_{a_j}, \mu_{a_j} \rangle 
\; \delta \left(\sum_{l=1}^m \mathbf{q}_{a_l}\right) 
d\mathbf{q}_{a_j} f_j (\mathbf{q}_{a_j}, 
\mu_{a_j})
\label{r.70}
\eeq
where $\vert f_\alpha \rangle$ denotes the product of square 
integrable functions $f_j(\mathbf{q}_{a_j},\mu_{a_j})$ of the 
momentum and spin of each bound cluster in the channel $\alpha$
and where we have factored out the total momentum conserving delta function.
In this notation the channel projectors (\ref{r.53}) can be expressed
as
\beq
\Pi_{\alpha} = \Phi_{\alpha}\Phi^{\dagger}_{\alpha}.
\label{r.71}
\eeq

The asymptotic Hilbert space for the reaction model is defined by
\beq
{\cal H}_{as,{\cal C}} := \oplus_{\alpha \in {\cal C} }
{\cal H}_{\alpha}.
\label{r.72}
\eeq
The sum of the $\Phi_{\alpha}$ defines a mapping from the asymptotic
Hilbert space to the model Hilbert space by
\beq
\Phi_{\cal C} := \sum_{\alpha \in {\cal C}} \Phi_{\alpha},
\label{r.73}
\eeq
were each $\Phi_{\alpha}$ is understood to act on the corresponding
channel subspace ${\cal H}_{\alpha}$.
Note that because of (\ref{r.62}) and (\ref{r.71}) the range of $\Phi_{\cal
  C}$ and $\Pi_{\cal C}$ coincide.

Symmetrized scattering channel wave functions are defined by the
strong limits
\beq
\vert \Psi^{\pm}_{\alpha} \rangle 
= \lim_{t \to \pm \infty} A e^{i M_\pi t}\Phi_\alpha e^{-i M_\alpha t} \vert
f_\alpha \rangle = 
\lim_{t \to \pm \infty} e^{i M_\pi t} A \Phi_\alpha e^{-i M_\alpha t} \vert
f_\alpha \rangle ,
\label{r.74}
\eeq
where $M_\alpha$ is the invariant mass of the asymptotic initial or
final state
\beq
M_\alpha =\sum_{j\in \alpha} \omega_{\lambda_{a_j}}\left(\mathbf{q_{a_j}}^2\right)
= \sum_{j \in \alpha} \sqrt{\lambda_{a_j}^2 + \mathbf{q}_{a_j}^2}  ,
\label{r.75}
\eeq
and the normalization of $\vert f_\alpha \rangle$ is chosen so
$\langle \Psi_\alpha^{\pm} \vert \Psi_\alpha^{\pm} \rangle =1$.  The
replacement of the Hamiltonian by the mass operator in (\ref{r.74}) is
justified \cite{fcwp:1982,bkwp} by the invariance principle
\cite{kato,cg2}.  Formally it corresponds to calculating the
Poincar\'e invariant $S$ matrix in the zero-momentum frame.

The relativistic $S$ matrix is defined for each initial and 
final channel $\beta, \alpha \in {\cal C}$ by  
\beq
S_{\alpha \beta}:= \langle \Psi^+_\alpha \vert \Psi^-_\beta \rangle = 
\lim_{t \to \infty} 
\langle f_\alpha \vert 
e^{i M_\alpha t} \Phi_\alpha^{\dagger}
A e^{-2i M_\pi t}A\Phi_\beta e^{i M_\beta t} \vert
f_\beta \rangle .
\label{r.76}
\eeq
Since $[M_{\pi},A]=0$ and $A^2=A$ one symmetrizer can be eliminated.
It is convenient to replace the initial and final states $\vert
f_{\alpha/\beta}\rangle $ by channel mass eigenstates with sharp
momenta $\vert \alpha/\beta \rangle $ and insert an $e^{-\epsilon t}$
factor to control the integral
\begin{eqnarray}
\langle \Psi^+_\alpha \vert \Psi^-_\beta \rangle & = & 
\langle \alpha \vert \Phi_\alpha^{\dagger} A \Phi_\beta \vert
\beta \rangle + 
\lim_{\epsilon\to 0^+}\int_0^\infty dt [{d \over dt}  
\langle \alpha \vert 
e^{i M_\alpha t} \Phi_\alpha^{\dagger}
A e^{-2i M_\pi t}A\Phi_\beta e^{i M_\beta t} \vert
\beta \rangle e^{-\epsilon t}] \cr 
 &=&\langle \alpha \vert \Phi_\alpha^{\dagger}
A\Phi_\beta \vert \beta \rangle 
- i \lim_{\epsilon\to 0^+} \int_0^\infty   
\langle \alpha \vert 
e^{i m_\alpha t} ( \Phi_\alpha^{\dagger}M_{\pi} - 
m_{\alpha} \Phi_\alpha^{\dagger}) 
e^{-2i M_\pi t}A\Phi_\beta e^{i m_\beta t} \vert
\beta \rangle e^{-\epsilon t} \nonumber \\
& & -i \lim_{\epsilon\to 0^+} \int_0^\infty   
\langle \alpha \vert 
e^{i m_\alpha t} \Phi_\alpha^{\dagger}
Ae^{-2i M_\pi t}
(M_{\pi} \Phi_\beta - \Phi_\beta m_\beta)  
e^{i m_\beta t} \vert
\beta \rangle e^{-\epsilon t} ,
\label{r.77} 
\end{eqnarray}
where it is understood that the limit is to be taken after smearing
with wave packets.  The same result would be obtained without
introducing the $\epsilon$ factor if the wave packets were retained.

The quantity $m_{\alpha}$ is the sharp-momentum eigenvalue of
$M_\alpha$ given by (\ref{r.75}), similarly for $m_{\beta}$.  It is
useful to introduce the average of the initial and final invariant
mass, defined by
\beq
\bar{m} = {1 \over 2} (m_{\alpha} + m_{\beta})
\label{r.78} 
\eeq
and note that 
\beq
\int_0^{\infty} e^{-2i (M_{\pi} - \bar{m} -i \epsilon)} =
\lim_{\epsilon \to 0^+} { i \over 2}{1 \over \bar{m} -M_{\pi} + i \epsilon}
:= { i \over 2}{1 \over \bar{m} -M_{\pi} + i 0^+}. 
\label{r.79} 
\eeq
Using (\ref{r.79}) in (\ref{r.77}) gives 
\begin{eqnarray}
\langle \alpha \vert 
\Phi_\alpha^{\dagger}
A \Phi_\beta \vert
\beta \rangle 
&+& {1\over 2}   \lim_{\epsilon \to 0^+} \langle \alpha \vert 
( \Phi_\alpha^{\dagger}M_{\pi} - 
m_{\alpha} \Phi_\alpha^{\dagger})
{ 1 \over \bar{m} - M_{\pi} + i\epsilon^+} 
A\Phi_\beta \vert
\beta \rangle   \nonumber \\
&+& {1 \over 2} \lim_{\epsilon \to 0^+}   
\langle \alpha \vert 
\Phi_\alpha^{\dagger}
A { 1 \over \bar{m} - M_{\pi} + i\epsilon^+} 
(M_{\pi} \Phi_\beta - \Phi_\beta m_\beta)  \vert
\beta \rangle .
\label{r.80} 
\end{eqnarray}

Applying the second resolvent identities as outlined in Appendix~\ref{appendixB}, 
the resulting expression for the approximate $S$ matrix element is:
\begin{eqnarray}
\langle \Psi^+_{\alpha} \vert \Psi^-_{\beta} \rangle& =& \langle \alpha_r \vert 
\Phi_\alpha^{\dagger} A \Phi_\beta \vert
\beta \rangle   \delta_{\alpha \beta}
- 2 \pi i \delta (m_\alpha - m_\beta) 
\Big[
\langle \alpha \vert  
\Phi_\alpha^{\dagger} A 
( M_{\pi} \Phi_\beta  - m_\beta \Phi_\beta )  \vert
\beta \rangle  \cr
&+& \langle \alpha \vert  
\left( \Phi_\alpha^{\dagger}M_{\pi} - 
m_{\alpha} \Phi_\alpha^{\dagger}\right)  
{ 1 \over m_{\beta} - M_{\pi} + i0^+}A
( M_{\pi} \Phi_\beta  - m_\beta \Phi_\beta )
\Big]
\beta \rangle .
\label{r.87} 
\end{eqnarray}
Note that $M$ and $\Pi$ normally have cluster expansions (see appendix
A after equation (\ref{a.65})) .  For a given partition $b$ of the particles into disjoint clusters
of the particles, $M_{b\pi}$ is obtained from $M_\pi$ by turning off
interactions between particles in different cluster of the partition
$b$ and eliminating channel projection operators in $\Pi$ that bind
particles in different cluster of $b$.  It follows that
\beq
(M_\pi - M_{b\pi}) \Phi_{\beta} = (M_\pi - m_{\beta})\Phi_{\beta}
\label{r.88} 
\eeq
because $\Phi_{\beta}$ is an eigenstate of $M_{b\pi}$ with eigenvalue
$m_{\beta}$ .
Defining
\beq
M^b_{\pi}:= M_\pi - M_{b\pi}
\label{r.89} 
\eeq
leads to the expression for the $S$-matrix elements in this
approximation
\begin{eqnarray}
\langle \Psi^+_{\alpha} \vert \Psi^-_{\beta} \rangle & =& 
\langle \alpha \vert \Phi_\alpha^{\dagger}
A \Phi_\beta \vert \beta \rangle  \delta_{\alpha \beta}  \cr  
&-& 2 \pi i \delta (m_\alpha - m_\beta) 
\Big[ \langle \alpha \vert  
\Phi_\alpha^{\dagger} A M_{\pi}^b  \Phi_\beta  \vert
\beta \rangle  + \langle \alpha \vert \Phi_\alpha^{\dagger}M^a_{\pi} 
{ 1 \over m_\beta - M_{\pi} + i0^+}A
M_{\pi}^b \Phi_\beta \vert \beta \rangle \Big] .
\label{r.90} 
\end{eqnarray}
The symmetrized approximate transition operator that acts on the open
channel spaces is
\beq
T_{\alpha \beta} := \Phi_\alpha^{\dagger} A \;
M_{\pi}^b  \Phi_\beta    
+   
\Phi_\alpha^{\dagger}M^a_{\pi} 
{ 1 \over m_\beta - M_{\pi} + i0^+}A
M_{\pi}^b \Phi_\beta
\label{r.91} .
\eeq
Note that in this form all of the internal degrees of freedom do not
appear in the transition matrix.  This is because the operators
$\Phi_\beta$ and $\Phi_\alpha^{\dagger}$ project the standard form of
the transition operators on the asymptotic channels subspaces.  The
result is that the internal degrees of freedom associated with the
bound clusters do not appear in $T_{\alpha \beta}$.

Both equation (\ref{r.90}) and (\ref{r.91}) contain an overall
momentum-conserving delta function that can be factored out of both
equations.

One would like to get integral equations directly for $T_{\alpha
\beta}$, which avoid having to treat all of the unphysical degrees
of freedom in the unprojected transition operators.  In order to
construct such equations we use (\ref{r.61}) and (\ref{r.71} ) to get the
following identity
\beq
\Pi_{\cal C}  = \sum_{\gamma \in {\cal C}} \Sigma^{\#} 
\Phi_\gamma \Phi_{\gamma}^{\dagger} . 
\label{r.92} 
\eeq
Inserting (\ref{r.92}) in the expression (\ref{r.91}) for $T$ gives the 
following expression for the projected transition operators
\beq
T_{\alpha \beta} :=
\Phi_\alpha^{\dagger} A
M_{\pi}^b  \Phi_\beta    
+   
\sum_{\gamma} \Phi_\alpha^{\dagger}M^a_{\pi} A
\Sigma^{\#} \Phi_\gamma \Phi_{\gamma}^{\dagger}
{ 1 \over m_\beta - M_{\pi} + i0^+} A
M_{\pi}^b \Phi_\beta .
\label{r.93} 
\eeq
Using the second resolvent identity from (\ref{r.94})
in (\ref{r.93}) gives

\begin{eqnarray}
T_{\alpha \beta} &=&
\Phi_\alpha^{\dagger} A M_{\pi}^b  \Phi_\beta  \cr  
& & +
\sum_{\gamma} \Phi_\alpha^{\dagger}M^a_{\pi} 
\Sigma^{\#} A \Phi_\gamma 
{ 1 \over m_\beta - m_\gamma + i0^+}
\left[\Phi_{\gamma}^{\dagger}A
M_{\pi}^b \Phi_\beta
+
\Phi_{\gamma}^{\dagger}
M^c_{\pi}
{ 1 \over m_\beta - M_{\pi} + i0^+}
A M_{\pi}^b \Phi_\beta \right] 
\label{r.95} 
\end{eqnarray}
which is an integral equation for $T_{\alpha \beta}$ 
\beq
T_{\alpha \beta} =  
\Phi_\alpha^{\dagger} A
M_{\pi}^b  \Phi_\beta    
+   
\sum_{\gamma\in {\cal C}} \Phi_\alpha^{\dagger} M^a_{\pi} 
\Sigma^{\#} A \Phi_\gamma 
{ 1 \over m_\beta - m_\gamma + i0^+}
T_{\gamma\beta}.
\label{r.96} 
\eeq
Here the sum is over all retained channels in $\gamma \in{\cal C}$. 

In general equation (\ref{r.96}) does not have a compact iterated
kernel which allows one to compute uniformly convergent
approximations.  It can be recast into such a form that the iterated
kernel is compact.  The basic idea is simple in principle, but the
operators are can be complicated depending on the reaction mechanism.

Abstractly expressed, equation (\ref{r.96}) has the form
\beq
T_{\alpha \beta} = D_{\alpha \beta} +\sum_{\gamma \in{\cal C}}
K_{\alpha \gamma}T_{\gamma \beta}.
\label{r.96a} 
\eeq
The kernel $K_{\alpha \gamma}$ has a cluster expansion.  For each
partition $c$ of the $N$ particle system into subsystems it can be
expressed as
\beq
K_{\alpha \gamma} = K_{c\alpha \gamma}+ K^c_{\alpha \gamma}
\label{r.96b} 
\eeq
where
$K_{\alpha \gamma}$ 
is the part of $K_{\alpha \gamma}$ that commutes with the
$\mathbf{q}_{ck}$ and $K^c_{\alpha \gamma}$ is the remainder.
For each partition $c$ we can construct 
\beq
(I - K_c)^{-1}_{\alpha\beta} .
\label{r.96e} 
\eeq
With this, for each partition $c$ the system of equations has the form
\beq
T_{\alpha \beta} = 
(I - K_c)^{-1}_{\alpha\delta}
D_{\delta \beta} +\sum_{\gamma \in{\cal C}}
(I - K_c)^{-1}_{\alpha\delta}K^c_{\delta \gamma} T_{\gamma \beta}.
\label{r.96f} 
\eeq
The following equation 
\beq
T_{\alpha \beta} = 
\sum_{c,n_c\geq 2} (-)^{n_c}(n_c-1)! (I - K_c)^{-1}_{\alpha\delta}
D_{\delta \beta} +\sum_{\gamma \in{\cal C}}
\sum_{c,n_c\geq 2} (-)^{n_c}(n_c-1)! 
(I - K_c)^{-1}_{\alpha\delta}K^c_{\delta \gamma}T_{\gamma \beta},
\label{r.96g} 
\eeq
where $n_c$ is the number of disjoint clusters in the partition $c$,
has a connected iterated kernel \cite{comb}. 
All of the terms in these equations only involve degrees of freedom
in the model Hilbert space.

In general the individual terms $(I - K_c)^{-1}_{\alpha\delta}$
have to be constructed recursively from subsystem equations,
however for reaction theories these operators are generally modeled.
Iterating these equations gives a generalization of the usual 
multiple scattering series \cite{watson,kmt}.


\section{Identical particles}

For systems of identical particles the number of channels in the 
scattering equations can be significantly reduced.  
For identical particles note that for each channel a permutation
operator either leaves the channel unchanged or transforms it to an
equivalent channel.  The permutations that leave the channel unchanged
involve permutation of particles in each asymptotic bound state, or
exchanges of identical asymptotic bound states.  There are
$n_{a_1}!\cdots n_{a_{m}}!$ permutations that leave each cluster of an
$m$ cluster channel $\gamma$ unchanged.  There are also $s!$ exchanges
for $s$ identical clusters with identical bound states.

Two channels that are related by permutation are called permutation
equivalent.  Those that are not are called permutation inequivalent.
Let $[\gamma]$ be the equivalence class of channels equivalent to
$\gamma$.  Let $n_{[\gamma]}$ be the number of channels in $[\gamma]$,
\beq
n_{[\gamma]} = {N! \over n_{a_1}!\cdots n_{a_{m}}! s_1! \cdot s_k!} .
\label{r.97}
\eeq

For each channel $\gamma$ the symmetrizer can be decomposed as follows
\beq
A = {1 \over N!} \sum P_{\sigma} = 
{1 \over n_{[\gamma]}} \sum_{\delta \in [\gamma]}
P_{\delta \gamma}A_{\gamma} =
{1 \over n_{[\delta]}} \sum_{\delta \in [\gamma]}
A_{\gamma} P_{\gamma \delta} 
\label{r.98} 
\eeq
where the permutation operator $P_{\sigma}$ is defined to include a
factor of $(-)^{\vert \sigma \vert}$ for identical fermions.  The
channel sum in the $T$-matrix equation can be decomposed into a sum
over equivalence classes of channels and a sum over elements in each
equivalence class
\beq
\sum_{\gamma \in {\cal C}} = 
\sum_{[\gamma] \in {\cal C}}\sum_{\gamma \in [\gamma]} .
\label{r.99} 
\eeq
Using this in the integral equation (\ref{r.96}) gives 
\beq
T_{\alpha \beta} =  
\Phi_\alpha^{\dagger} 
{1 \over n_{[\alpha]}} \sum_{\delta \in [\alpha]}
P_{\alpha \delta} 
M_{\pi}^b  \Phi_\beta    
+   
\sum_{[\gamma] \in {\cal C}} \sum_{\gamma \in [\gamma]} 
\Phi_\alpha^{\dagger} M^a_{\pi} 
\Sigma^{\#} {1 \over n_{[\gamma]}} \sum_{\delta \in [\gamma]} 
P_{\delta \gamma}\Phi_\gamma 
{ 1 \over m_\beta - m_\gamma + i0^+}
T_{\gamma\beta} .
\label{r.100} 
\eeq
We note that 
\[
\sum_{\gamma \in [\gamma]}
{1 \over n_{[\gamma]}}
\sum_{\delta \in [\gamma]}
P_{\delta \gamma}\Phi_\gamma 
{ 1 \over m_\beta - m_\gamma + i0^+}
T_{\gamma\beta} =
\]
\beq
\sum_{\delta \in [\gamma]}
P_{\delta \gamma}\Phi_\gamma 
{ 1 \over m_\beta - m_\gamma + i0^+}
T_{\gamma\beta} 
\label{r.101} 
\eeq
which when used in (\ref{r.100}) gives the symmetrized equation
\beq
T_{\alpha \beta} =  
\Phi_\alpha^{\dagger}
{1 \over n_{[\alpha]}} \sum_{\delta \in [\alpha]}
P_{\alpha \delta}  
M_{\pi}^b  \Phi_\beta    
+   
\sum_{[\gamma] \in {\cal C}} \sum_{\delta \in [\gamma]} 
\Phi_\alpha^{\dagger} M^a_{\pi} 
\Sigma^{\#}  
P_{\delta \gamma}\Phi_\gamma 
{ 1 \over m_\beta - m_\gamma + i0^+} 
T_{\gamma\beta}.
\label{r.102} 
\eeq
In this equation $\gamma$, $\alpha$ and $\beta$ are arbitrary but
fixed elements of the classes $[\gamma]$, $[\alpha]$ and $[\beta]$.

The effective interactions for this symmetrized equation are
\beq
\sum_{\delta \in [\gamma]} 
\Phi_\alpha^{\dagger} M^a_{\pi} 
\Sigma^{\#}  
P_{\delta \gamma}\Phi_\gamma .
\label{r.102b}
\eeq
The kernel of this equation is only compact for models with only two
cluster channels.  When the reaction mechanism includes channels with
three or more clusters then it is necessary to construct an equivalent
compact kernel equation or to establish that there are no non-zero
solutions to the homogeneous equations.

These equations give the approximate transition operator derived in
section 6 however they do not include the effects of the eliminated
channels.  We could have replaced $M_\Pi$ by
\beq
M_\Pi \to M_{\Pi} + \Pi M \Pi' (\lambda - \Pi' M \Pi' +i0 )^{-1}  
\Pi'M \Pi
\eeq
with $I = \Pi + \Pi'$, which would lead to equations of the same form
with the interaction terms replaced by energy dependent optical
potentials.  Since this decomposition still preserves the rotational
invariance, it will lead to irreducible representations of the
Poincar\'e group.


\section{(d,p) Reactions}

To illustrate the formalism we consider the case of a (d,p) reaction.
We choose the dominant reaction channels ${\cal C}$ to include (1) the
deuteron and an $A$-particle target nucleus, (2) the deuteron and an
$A$-particle excited nucleus, $A^*$, (3) two nucleons and the target nucleus,
and (4) a nucleon an $A+1$ particle nucleus, and all channels
generated by exchange of identical nucleons.  Here we treat the protons
and neutrons as different isospin states of a nucleon.  
This leads to an effective three-body problem.
For low energy (d,p) reactions this approach was pioneered in
Refs.~\cite{Deltuva:2009fp,Alt:2007wm,Mukhamedzhanov:2012qv} in
the framework of the Faddeev AGS equations. 
Within a Poincar\'e invariant formulation the dynamical equations governing 
this system are formally given 
by (\ref{r.102}).
The channel injection operators are 
\begin{eqnarray}
\Phi_1 &:=& \vert \mathbf{P} ; (m_d,{1}) \mathbf{q}_d , \mu_d ,
(m_A,j_A) -\mathbf{q}_d , \mu_A \rangle  
\label{r.103} \\
\Phi_2& :=& \vert \mathbf{P} ; (m_d,{1}) \mathbf{q}_d , \mu_d ,
(m_{A^*},j_{A^*}) -\mathbf{q}_d , \mu_{A^*} \rangle  
\label{r.104} \\
\Phi_3& :=& \vert \mathbf{P} ; (m_N,{1 \over 2}) \mathbf{q}_N , \mu_N ,
(m_N,{1 \over 2}) \mathbf{q}_N' , \mu_N' 
(m_{A},j_{A}) -\mathbf{q}_N-\mathbf{q}_N'  , \mu_{A} \rangle  
\label{r.105} \\
\Phi_4& :=& \vert \mathbf{P} ; (m_N,{1 \over 2}) \mathbf{q}_N , \mu_N ,
(m_{A+1},j_{A+1}) -\mathbf{q}_N , \mu_{A+1} \rangle .  
\label{r.106}
\end{eqnarray} \\
The full set of channels ${\cal C}$ is generated by applying 
permutations to these channels.

The operator $\Sigma$ is given by
\beq
\Sigma := \sum_{\gamma \in [1]} 
P_{\gamma 1} \Phi_1 \Phi_1^{\dagger} P_{1 \gamma }^{\dagger} +
\sum_{\gamma \in [2]} 
P_{\gamma 2} \Phi_2 \Phi_2^{\dagger} P_{2 \gamma }^{\dagger} +
\sum_{\gamma \in [3]} 
P_{\gamma 3} \Phi_3 \Phi_3^{\dagger} P_{3 \gamma }^{\dagger} +
\sum_{\gamma \in [4]} 
P_{\gamma 4} \Phi_4 \Phi_4 P_{4 \gamma }^{\dagger}
\label{r.107}
\eeq
and  
\beq
\Pi_{\cal C} := \Sigma^{\#} \Sigma .
\label{r.108}
\eeq
The model mass operator is
\beq
M_{\Pi} = \Pi_{\cal C}  M \Pi_{\cal C} .
\label{r.109}
\eeq
The individual channel masses are 
\begin{eqnarray}
m_1& =& \sqrt{m_d^2 + \mathbf{q}_d^2} + \sqrt{m_A^2 + \mathbf{q}_d^2} 
\label{r.110} \nonumber \\
m_2& =& \sqrt{m_d^2 + \mathbf{q}_d^2} + \sqrt{m_{A^*}^2 + \mathbf{q}_d^2}
\label{r.111} \nonumber \\
m_3& =& \sqrt{m_N^2 + \mathbf{q}_{N1}^2} + \sqrt{m_N^2 + \mathbf{q}_{N2}^2}
+ \sqrt{m_{A}^2 + (\mathbf{q}_{N1}+\mathbf{q}_{N2})^2}
\label{r.112} \nonumber \\
m_4& =& \sqrt{m_N^2 + \mathbf{q}_N^2} + \sqrt{m_{A+1}^2 + \mathbf{q}_N^2} 
\label{r.113}
\end{eqnarray}
and 
\begin{eqnarray}
M_{\Pi i} &:=&  \Phi_i m_i \Phi_i^{\dagger}    
\label{r.114} \nonumber \\
M^i_{\Pi} &:=& M_{\Pi}- M_{\Pi i} .
\label{r.115}
\end{eqnarray}
The projected transition matrix elements are  
\begin{eqnarray}
T_{11}& =& 
\langle  \mathbf{P}  (m_d,{1}) \mathbf{q}_d , \mu_d ,
(m_A,j_A) -\mathbf{q}_N , \mu_A \vert 
T^{11} (z) \vert 
\mathbf{P}'  (m_d,{1}) \mathbf{q}'_d , \mu'_d ,
(m_A,j_A) -\mathbf{q}'_d , \mu'_A \rangle  \cr
&=&\delta (\mathbf{P} - \mathbf{P}')
t_{11} (\mathbf{q}_d , \mu_d,\mu_A;
\mathbf{q}_d' , \mu_d',\mu_A',z )
\label{r.116} \\
T_{21} &=& 
\langle  \mathbf{P}  (m_d,{1}) \mathbf{q}_d , \mu_d ,
(m_{A^*},j_{A^*}) -\mathbf{q}_d , \mu_{A^*} \vert 
T^{21} (z) \vert 
\mathbf{P}'  (m_d,{1}) \mathbf{q}'_d , \mu'_d ,
(m_A,j_A) -\mathbf{q}'_d , \mu'_A \rangle  \cr
&=&\delta (\mathbf{P} - \mathbf{P}')
t_{21} (\mathbf{q}_d , \mu_d,\mu_B;
\mathbf{q}_d' , \mu_d',\mu_A',z )
\label{r.117} \\
T_{31} &=& 
\langle  \mathbf{P}  (m_N,{1 \over 2}) \mathbf{q}_{N1} , \mu_{N1} ,
(m_N,{1 \over 2}) \mathbf{q}_{N2} , \mu_{N2}
(m_{A},j_{A}) -(\mathbf{q}_{N1}+\mathbf{q}_{N2}) , \mu_{A} \vert 
T^{31} (z)  \cr
&& \times 
\vert 
\mathbf{P}'  (m_d,{1}) \mathbf{q}'_d , \mu'_d ,
(m_A,j_A) -\mathbf{q}'_d , \mu'_A \rangle  \cr
&=&\delta (\mathbf{P} - \mathbf{P}')
t_{31} (\mathbf{q}_{N1} , \mu_{N1}, \mathbf{q}_{N2}, \mu_{N2},\mu_{A};
\mathbf{q}_d' , \mu_d',\mu_A',z )
\label{r.118} \\
T_{41} &=& 
\langle  \mathbf{P}  (m_N,{1\over 2}) \mathbf{q}_N , \mu_N ,
(m_{A+1},j_{A+1}) -\mathbf{q}_N , \mu_{A+1} \vert 
T^{41} (z) \vert 
\mathbf{P}'  (m_d,{1}) \mathbf{q}'_d , \mu'_d ,
(m_A,j_A) -\mathbf{q}'_d , \mu'_A \rangle  \cr
&=&\delta (\mathbf{P} - \mathbf{P}')
t_{41} (\mathbf{q}_N , \mu_N,\mu_{A+1};
\mathbf{q}_d' , \mu_d',\mu_A',z ) 
\label{r.119}
\end{eqnarray}
where $z=\sqrt{m_d^2+\mathbf{q}_d^{\prime 2}}+
\sqrt{m_A^2+\mathbf{q}_A^{\prime 2}}+i0^+$ is the 
incident invariant energy. 
For a reasonable sized target nucleus the input to the
equations, while well defined, must ultimately be treated 
phenomenologically.  These elements are interactions and kernel terms.
The $11$ driving term is 
\begin{eqnarray}
V_{11} 
&= & {1 \over N_{[1]}} \sum_{\gamma \in [1]}
 \langle  \mathbf{P}  (m_d,{1}) \mathbf{q}_d , \mu_d ,
(m_A,j_A) -\mathbf{q}_d , \mu_A \vert 
P_{1 \gamma} M_{\Pi}^1 \vert 
\mathbf{P}'  (m_d,{1}) \mathbf{q}'_d , \mu'_d ,
(m_A,j_A) -\mathbf{q}'_d , \mu'_A \rangle  \cr
&=& \delta (\mathbf{P} - \mathbf{P}')
v_{11} (\mathbf{q}_d , \mu_d,\mu_A;
\mathbf{q}_d' , \mu_N',\mu_A')
\label{r.120}
\end{eqnarray}
where 
\beq
v_{11} (\mathbf{q}_d , \mu_d,\mu_A;
\mathbf{q}_d' , \mu_d',\mu_A')
\label{r.121}
\eeq
which is a rotationally invariant functions of the $\mathbf{q}_i$ and
constituent spins.  There are three other driving terms, $V_{21}$,
$V_{31}$, $V_{41}$ associated with the three other final channels.

The interaction part of the kernel has 16 terms of the form 
$K_{ij}$.  They have a form similar to $K_{11}$, which is given by   
\begin{eqnarray}
K_{11}& =& 
\sum_{\gamma \in [1]}
\langle  \mathbf{P}  \left(m_n,{1}\right) \mathbf{q}_d , \mu_d ,
(m_A,j_A) -\mathbf{q}_d , \mu_A \vert 
M_{\Pi}^1 \Sigma^{\#}P_{\gamma 1} \vert
\mathbf{P}'  \left(m_d,{1}\right) \mathbf{q}'_d , \mu'_d ,
(m_A,j_A) -\mathbf{q}'_d , \mu'_A \rangle  \cr
& &
=\delta (\mathbf{P} - \mathbf{P}')
k_{11} (\mathbf{q}_d , \mu_d,\mu_A;
\mathbf{q}_d' , \mu_d',\mu_A')
\label{r.122}
\end{eqnarray}
where 
\beq
k_{11} (\mathbf{q}_d , \mu_d,\mu_A;
\mathbf{q}_d' , \mu_d',\mu_A')
\label{r.123}
\eeq
is a rotationally invariant kernel.

The integral equation is a four by four matrix of equations involving
all four amplitudes.  After factoring out the overall momentum
conserving delta function we get
\begin{eqnarray}
\lefteqn {t_{11} (\mathbf{q}_d , \mu_d,\mu_A;
\mathbf{q}_d' , \mu_d',\mu_A',m_{A'} ) } \cr
&= & 
v_{11} (\mathbf{q}_d , \mu_d,\mu_A;
\mathbf{q}_d' , \mu_d',\mu_A')   \cr 
&+&\sum_{\mu_d'',\mu_A''} \int {k_{11} (\mathbf{q}_d , \mu_d,\mu_A;
\mathbf{q}_d'' , \mu_d'',\mu_A'') d\mathbf{q}_d''
t_{11} (\mathbf{q}_d'' , \mu_d'',\mu_A'';
\mathbf{q}_d' , \mu_d',\mu_A',m_{A'} ) \over 
m_{1'} - m_1'' +i 0^+} \cr
&+&\sum_{\mu_d'',\mu_{A^*}''} \int {k_{12} (\mathbf{q}_d , \mu_d,\mu_A;
\mathbf{q}_d'' , \mu_d'',\mu_A^{*\prime \prime}) d\mathbf{q}_d''
t_{21} (\mathbf{q}_d'' , \mu_d'',\mu_A^{*\prime\prime};
\mathbf{q}_d' , \mu_d',\mu_A',m_{A'} ) \over 
m_{1'} - m_2'' +i 0^+}  \cr
&+&\sum_{\mu_{N1}'',\mu_{N2}'',\mu_A''} \int d\mathbf{q}_{N_1}''d\mathbf{q}_{N_1}'' \times
\cr
&&  {k_{13} (\mathbf{q}_d , \mu_d,\mu_A;
\mathbf{q}_{N1}'' , \mu_{N1}'', \mathbf{q}_{N2}'', 
\mu_{N2}'',\mu_{A}'') 
t_{31} ( \mathbf{q}_{N1}'' , \mu_{N1}'', \mathbf{q}_{N2}'', 
\mu_{N2}'',\mu_{A}'';
\mathbf{q}_d' , \mu_d',\mu_A',m_{A'} ) \over 
m_{1'} - m_3'' +i 0^+} \cr
&+& \sum_{\mu_{N}'',\mu_{A+1}''} \int
{k_{14} (\mathbf{q}_d , \mu_d,\mu_A;
\mathbf{q}_N'' , \mu_N'',\mu_{A+1}'') d\mathbf{q}_N''
t_{41} (\mathbf{q}_N'' , \mu_N'',\mu_{A+1}'';
\mathbf{q}_d' , \mu_d',\mu_A',m_{A'} ) \over 
m_{1'} - m_4'' +i 0^+} . 
\label{r.124}
\end{eqnarray}
This is the first of four coupled equations, the others are for $t_{21}$,
$t_{31}$, $t_{41}$.  These equations have the same general structure.

These are a set of four coupled channel equations for the four
symmetrized transition matrix elements.  The kernel has disconnected
terms which remain disconnected upon iteration.  These can be replaced
by equivalent connected kernel equations using the methods discussed
at the end of section 6.

In this case a direct solution is easier.  The starting point is
equations (\ref{r.124} $\cdots$ ) which have the abstract form:
\beq 
t_{i1} = v_{i1} + 
\sum_{j=1}^4  K_{ij} t_{j1}
\label{r.125}
\eeq
The first step is to eliminate breakup amplitude $(j=3)$ using
\beq 
t_{31} = (1-K)_{33}^{-1}v_{31} + 
(1-K)_{33}^{-1} K_{31} t_{11} +
(1-K)_{33}^{-1} K_{32} t_{21} +
(1-K)_{33}^{-1} K_{34} t_{41} .
\label{r.126}
\eeq
The second step is to insert this into the remaining three equations
\begin{eqnarray}
t_{k1}& =& v_{k1} +
K_{k3}(1-K)_{33}^{-1}v_{31}  \cr
&+&(K_{k1} + K_{k3}(1-K)_{33}^{-1} K_{31}) t_{11}+
(K_{k2} + K_{k3}(1-K)_{33}^{-1} K_{32}) t_{21} \cr
&+&(K_{k4} + K_{k3}(1-K)_{33}^{-1} K_{34}) t_{41} .
\label{r.127}
\end{eqnarray}
The last step is to make the kernels connected upon iteration
which gives the following three coupled equations for the
two-cluster amplitudes:
\begin{eqnarray}
t_{k1}& =& (I - K_{kk} + K_{k3}(1-K)_{33}^{-1} K_{3k})^{-1} v_{k1} 
K_{k3}(1-K)_{33}^{-1}v_{31} \cr
&+&\sum_{l\not= k,3}(I - K_{kk} + K_{k3}(1-K)_{33}^{-1} K_{3k})^{-1}
(K_{kl} + K_{k3}(1-K)_{33}^{-1} K_{3l})^{-1} t_{l1} .
\label{r.128}
\end{eqnarray}
These equations can be solved using Faddeev methods.  The breakup
amplitude can be calculated from these solutions using (\ref{r.126}).  The
effective interactions are complicated many-body operators that, while
precisely defined, have to be modeled in practice.  The interactions
include both effective two and three-body interactions.  In this model
the ``three-body forces'' will be important because they include
effects from the exchange channels.  If one wants to include corrections
from some of the eliminated channels, then the interactions are
replaced by energy-dependent optical potentials.

The number of continuous variables is the same as one would get on a
three-body Faddeev equation.  Unlike the relativistic few-body
problem, depending on the charge of the core, Coulomb effects may have
to be included.  This requires an additional analysis due to the 
long-range nature of the Coulomb interaction.


\section{Summary}

In the preceding sections a formulation of a theory for nuclear reactions is given
in a representation of Poincar\'e invariant quantum mechanics where the interactions
are invariant with respect to kinematic translations and rotations. 
It has the advantage that the framework is valid
for any number of particles and the dynamical equations have the same
number of variables as the corresponding non-relativistic equations.
We discussed the approximations that emphasize the dominant degrees of freedom so
that both unitarity and exact Poincar\'e invariance are preserved.
Poincar\'e invariance is an exact symmetry that is realized by a
unitary representation of the Poincar\'e group on the corresponding 
Hilbert space.  The dynamics is generated by a Hamiltonian.  This
feature is shared with the Galilean invariant formulation of
non-relativistic quantum mechanics. The Hamiltonian of the
corresponding relativistic formulation differs in how the two-body
interactions are embedded in the  Hamiltonian (mass operator).

As specific example of the formulation we considered the case of
(d,p) reactions, which leads to an effective three-body problem and
worked out the relevant transition matrices between the different
channels. Similar to the non-relativistic Faddeev
equations~\cite{Mukhamedzhanov:2012qv,Deltuva:2013jna}, the Poincar\'e
invariant formulation allows the explicit inclusion of target
excitations as additional channel.

Though a practical implementation is not yet in sight, having a
theoretical framework that allows one to isolate the dynamics
associated with a given set of reaction channels at relativistic
energies, and systematically compute corrections, provides precise
definitions of the quantities that must be modeled in applications.
Specifically, as experimental capabilities in investigating reaction
with rare isotopes are continuously refined, the assumptions and
approximations use to study reactions at higher energies need to
be examined as approximations to a relativistic theory of reactions.


\appendix

\section{Moore-Penrose generalized inverse}
\label{appendixA}

In this appendix we discuss methods for computing the Moore Penrose
generalized inverse.
The definition 
\beq
\Sigma^{\#}_{\alpha} :=  \Pi_\alpha \Sigma_{\cal C}^{\#}   
\label{a.57}
\eeq
implies
\beq
\Pi_{\cal C} = \sum_{\alpha \in {\cal C}} \Sigma^{\#}_{\alpha}.
\label{a.58}
\eeq
Multiplying both sides of (\ref{a.57}) by $\Pi_\alpha$ and rearranging
terms gives
\beq
\Sigma^{\#}_{\alpha} = \Pi_{\alpha} - \sum_{\beta \not= \alpha \in {\cal C}} 
\Pi_{\alpha} \Sigma^{\#}_{\beta} .
\label{a.59}
\eeq
For two-cluster channels this set of equations, after factoring our
the total momentum-conserving delta functions, has a non-singular
compact iterated kernel, which can be uniformly approximated by a
finite-dimensional matrix.  This gives a straightforward means to
construct the solution to these equations using uniform
approximations.

The solution of (\ref{a.59}) can be used to calculate
\beq
\Pi_{\cal C} = \sum_{\alpha \in {\cal C}}  \Sigma^{\#}_{\alpha}.
\label{a.60}
\eeq
When the projectors in $\Sigma_{\cal C}$ include more than two
clusters channels the series and the non-zero eigenvalues of
$\Sigma_{\cal C}$ are bounded above zero then
\beq
\Pi_{\cal C} = \sum_{n=0}^{\infty} (1- \gamma \Sigma_{\cal C})^n 
\gamma \Sigma_{\cal C} = 
\gamma \Sigma_{\cal C} \sum_{n=0}^{\infty} (1- \gamma \Sigma_{\cal C})^n  
\label{a.61}
\eeq
will converge uniformly for $\gamma$ less that $1/$(number of channels).
The relevant iteration is 
\beq
\Pi (0) :=  \gamma \Sigma_{\cal C}
\label{a.62}
\eeq
\beq
\Pi (n+1) = \Pi (n)(1-\gamma \Sigma_{\cal C})
\label{a.63}
\eeq
\beq
\Pi_{\cal C}  = \lim_{n\to \infty} \Pi(n) . 
\label{a.64}
\eeq
The rate of convergence depends on both the choice of $\gamma$ and the
size of the smallest non-zero eigenvalue of $\Sigma_{\cal C}$.  There
is also a similar series for
\beq
\Sigma_\alpha^{\#}  = 
\gamma \Pi_\alpha  \sum_{n=0}^{\infty} (1- \gamma \Sigma_{\cal C})^n .  
\label{a.65}
\eeq
Cluster expansions for $\Sigma_\alpha^{\#}$ and $\Pi_{\cal C}$ can be developed 
from this representation.
 
An alternative way to calculate $\Sigma^{\#}\Phi_\alpha$, which uses
connected kernel equations, is based on the observation that the
resolvent of $X$ satisfies the Weinberg-Van Winter equations\cite{comb}
\[
{1 \over z-\Sigma} = \sum_{a,n_a\geq 2} C_a {1 \over z - \Sigma_a}
+ \sum_{a,n_a\geq 2} C_a (\Sigma-\Sigma_a) {1 \over z-\Sigma}
\]
where $\Sigma_a$ is the sum of all projectors that commute with translations of
the cluster of the partition $a$.  The coefficients 
$C_a$ are  
\beq
C_a = (-)^{n_a} (n_a-1)!
\eeq
where $n_a$ is the number of non-empty clusters in the partition $a$.
These equations always have compact kernels.  They can be solved
recursively (n the number of particles) to build up the ${1 \over z -
  \Sigma_a}$ that are the input to these equations.  The starting
point corresponds to the finest partitions where the resolvents have
the trivial form
\beq
{1 \over z - \Pi_\alpha} = \Pi_\alpha {1 \over z-1} + \Pi {1 \over z}
\eeq
This gives a Faddeev type of construction to find $\Pi$.  It requires
that the Moore-Penrose generalized inverse is bounded or equivalently
that the spectrum of $\Sigma$ has a gap between 0 and its first
non-zero eigenvalue.

The operator $\Sigma^{\#} \Phi_\alpha$ which appears in the integral
equation can be calculated using
\[
\Sigma^{\#} \Phi_\alpha =
\lim_{z \to 0} {1 \over X -z} \Phi_\alpha
\]

This limit makes sense because the range of $\Phi_\alpha$ is in the
range of $\Sigma$.  The Weinberg-Van Winter equation can be replaced
by
\beq
{1 \over z-\Sigma} \Phi_{\alpha} = \sum_{a,n_a\geq 2} C_a {1 \over z - \Sigma_a}
\Phi_{\alpha}
+ \sum_{a,n_a\geq 2} C_a (\Sigma-\Sigma_a) {1 \over z-\Sigma} 
\Phi_{\alpha}
\eeq


\section{Formulations with Resolvent Identities}
\label{appendixB}

The second resolvent identities are used in (\ref{r.80})   to obtain
\begin{eqnarray}
\lefteqn{\langle \alpha \vert \Phi_\alpha^{\dagger}
A \Phi_\beta \vert \beta \rangle } \cr
&+& {1\over 2}   \langle \alpha \vert ( \Phi_\alpha^{\dagger}M_{\pi} - 
m_{\alpha} \Phi_\alpha^{\dagger})\; A ( \Phi_\beta +  
{ 1 \over \bar{m} - M_{\pi} + i\epsilon^+} 
(M_\pi \Phi_\beta - \Phi_\beta m_\beta ))   
{ 1 \over \bar{m} - m_{\beta} + i\epsilon^+}
\vert \beta \rangle \cr 
&+& {1 \over 2}    
\langle \alpha \vert 
{ 1 \over \bar{m} - m_{\alpha} + i\epsilon^+}
(\Phi_\alpha^{\dagger} + (\Phi_\alpha^{\dagger} M_{\pi} - m_{\alpha} \Phi_\alpha^{\dagger})  { 1 \over \bar{m} - M_{\pi} + i\epsilon^+}) 
A(M_{\pi} \Phi_\beta - \Phi_\beta m_\beta)  \vert
\beta \rangle .
\label{r.81} 
\end{eqnarray}
Separating the kinematical and dynamical terms gives 
\begin{eqnarray}
\lefteqn{\langle \alpha \vert \Phi_\alpha^{\dagger}
A \Phi_\beta \vert \beta \rangle} \cr
&+& {1 \over 
m_\alpha - m_\beta + i\epsilon}
\langle \alpha_r \vert ( \Phi_\alpha^{\dagger}M_{\pi} - 
(m_{\alpha}-m_\beta +m_\beta) \Phi_\alpha^{\dagger}) A \Phi_\beta 
\vert \beta \rangle  \cr
&-& {1 \over m_\alpha - m_\beta - i\epsilon}
\langle \alpha_r \vert 
\Phi_\alpha^{\dagger}A ( M_{\pi} \Phi_\beta  - m_\beta \Phi_\beta ) 
\vert \beta \rangle  \cr
&+ & \langle \alpha \vert  ( \Phi_\alpha^{\dagger}M_{\pi} - 
m_{\alpha} \Phi_\alpha^{\dagger}) A 
{ 1 \over \bar{m} - M_{\pi} + i0^+}
( M_{\pi} \Phi_\beta  - m_\beta \Phi_\beta ) 
\vert
\beta \rangle 
\times \cr
& & 
\left[ {1 \over m_\alpha - m_\beta + i\epsilon}-
{1 \over m_\alpha - m_\beta - i\epsilon}\right].
\label{r.82} 
\end{eqnarray}
This becomes
\begin{eqnarray}
\lefteqn{\langle \alpha \vert \Phi_\alpha^{\dagger}
A \Phi_\beta \vert \beta_r \rangle } \cr
&+&\langle \alpha \vert  \Phi_\alpha^{\dagger} A(M_{\pi} \Phi_\beta 
- \Phi_\beta m_\beta) \vert \beta \rangle 
\left[ {1 \over m_\alpha - m_\beta + i\epsilon}-
{1 \over m_\alpha - m_\beta - i\epsilon} \right] \cr 
&+ & \langle \alpha \vert  \Phi_\alpha^{\dagger} A \Phi_\beta \vert
\beta \rangle {m_\beta - m_\alpha \over 
m_\alpha - m_\beta + i\epsilon} \times \cr
& &\langle \alpha \vert   ( \Phi_\alpha^{\dagger}M_{\pi} - 
m_{\alpha} \Phi_\alpha^{\dagger}) A 
{ 1 \over \bar{m} - M_{\pi} + i0^+}
( M_{\pi} \Phi_\beta  - m_\beta \Phi_\beta )
\vert \beta \rangle
\times \cr
& & 
\left[ {1 \over m_\alpha - m_\beta + i\epsilon}-
{1 \over m_\alpha - m_\beta - i\epsilon}\right]
\label{r.83} 
\end{eqnarray}
which is equal to 
\begin{eqnarray}
\lefteqn{\langle \alpha \vert \Phi_\alpha^{\dagger}
A \Phi_\beta \vert
\beta \rangle { i \epsilon \over m_\alpha - m_\beta +i\epsilon^+} } \cr
&+&\langle \alpha \vert  \Phi_\alpha^{\dagger} A(M_{\pi} \Phi_\beta 
- \Phi_\beta m_\beta) \vert \beta \rangle 
\left[ {1 \over m_\alpha - m_\beta + i\epsilon}-
{1 \over m_\alpha - m_\beta - i\epsilon} \right] \cr
& +&
\langle \alpha \vert  ( \Phi_\alpha^{\dagger}M_{\pi} - 
m_{\alpha} \Phi_\alpha^{\dagger}) A { 1 \over \bar{m} - M_{\pi} + i0^+}
( M_{\pi} \Phi_\beta  - m_\beta \Phi_\beta )
\vert \beta \rangle \times \cr
& & 
\left[ {1 \over m_\alpha - m_\beta + i\epsilon}-
{1 \over m_\alpha - m_\beta - i\epsilon}\right].
\label{r.84} 
\end{eqnarray}
The $\epsilon$ factors become
\beq
{i \epsilon \over m_\alpha -m_\beta +i\epsilon} = 
\delta_{\alpha \beta}
\label{r.85} 
\eeq
and
\beq
\left[ {1 \over 
m_\alpha - m_\beta + i\epsilon}-
{1 \over m_\alpha - m_\beta - i\epsilon}\right] =
{-2i\epsilon \over (m_\alpha-m_\beta)^2 + \epsilon^2} \to
-2 \pi i \delta (m_\alpha - m_\beta) .
\label{r.86} 
\eeq
The first term vanishes if $m_\alpha \not = m_\beta$ as $\epsilon \to 0$; 
it becomes 1 when the channels are the same - as a Kronecker
delta.

In order to obtain (\ref{r.95}) we 
note that $A^2=A$ has been used to put $A$ in two places separated by
operators that commute with $A$.  
Next the second resolvent equations
are used to arrive at
\begin{eqnarray}
\lefteqn{ { 1 \over m_\beta - M_{\pi} + i0^+} =} \cr
 { 1 \over m_\beta - M_{c\pi} + i0^+} &+ &
{ 1 \over m_\beta - M_{c\pi} + i0^+}
M^c_{\pi}
{ 1 \over m_\beta - M_{\pi} + i0^+} \cr
& & { 1 \over m_\beta - M_{c\pi} + i0^+}(1+
M^c_{\pi}
{ 1 \over m_\beta - M_{\pi} + i0^+}).
\label{r.94} 
\end{eqnarray}

\begin{acknowledgments}
This work was performed under the auspices of the U. S. Department of Energy,
Office of Nuclear Physics, under contract No. DE-FG02-86ER40286 with the
University of Iowa and No. DE-FG02-93ER40756 with Ohio University.
The authors thank R.C. Johnson for his invitation to embark in this work.
\end{acknowledgments}


\end{document}